\documentclass{article}

\usepackage{graphicx}
\usepackage{amsfonts}
\usepackage{amssymb}
\usepackage{amsmath}

\makeatletter
\@addtoreset{equation}{section}

\makeatother

\newlength{\GraphicsWidth}
\newcommand{\arctanh}{\mathop{\text{arctanh}}}
\newcommand{\exc}{\mathrm{exc}}
\renewcommand{\r}{\mathbf{r}}
\newcommand{\R}{\mathbf{R}}
\newcommand{\dslash}{\hskip.1ex\hbox to 0pt{/\hss}\kern-.1ex\partial}

\title{Two-dimensional two-component plasma with adsorbing impurities}

\author{Alejandro Ferrero\footnote{e-mail:al-ferre@uniandes.edu.co}
\ and Gabriel T\'ellez\footnote{e-mail:gtellez@uniandes.edu.co}\\ 
Departamento de
F\'{\i}sica\\
 Universidad de los Andes\\
A.A. 4976, Bogot\'a, Colombia
} 
\date{}

\begin{document}

\maketitle

\begin{abstract}
We study the behavior of the two-dimensional two-component plasma in
the presence of some adsorbing impurities. Using a solvable model, we
find analytic expressions for the thermodynamic properties of the
plasma such as the $n$-body densities, the grand potential, and the
pressure. We specialize in the case where there are one or two
adsorbing point impurities in the plasma, and in the case where there
are one or two parallel adsorbing lines. In the former case we study
the effective interaction between the impurities, due to the charge
redistribution around them. The latter case is a model for electrodes
with adsorbing sticky sites on their surface.
\end{abstract}


\section{Introduction}
A two-component plasma (TCP) is a system composed of two different
types of particles with charges $\pm e$ interacting through a Coulomb
potential. In this work, we study a two-dimensional model which is
exactly solvable. The solution of Poisson equation shows that in two
dimensions, the Coulomb interaction between two particles, with
charges $q_1$, $q_2$, at a distance $r$ from each other, is
\begin{equation}
\label{eq:Coulomb-pot}
v(r)=-q_1q_2\ln(r/L)  \,,
\end{equation}
where $L$ is an irrelevant length scale fixing the potential
reference.

Let us define the dimensionless coupling constant $\Gamma=\beta e^2$,
where $\beta=(k_B T)^{-1}$, $T$ the temperature and $k_B$ the
Boltzmann constant. The coupling constant $\Gamma$ is the ratio
between the electrostatic and thermal energy of the particles of the
gas. If $\Gamma<2$, the thermal energy is high enough to prevent the
collapse of particles of different sign. If $\Gamma\ge 2$, the system
is unstable against the collapse and the point-particles must be
replaced by hard disks of diameter~$a$. Using the analogy between the
two-component plasma and the sine-Gordon quantum field theory, exact
results for the thermodynamic properties of the Coulomb gas, in the
bulk~\cite{ST} and near a planar
interface~\cite{SJ-tcp-metal,Samaj-tcp-diele}, have been obtained when
$\Gamma<2$. For $\Gamma=2$, the classical two-dimensional
two-component plasma is equivalent to a quantum free Fermi
gas~\cite{uno}. This fact allows us to find analytic expressions for the
thermodynamic properties and correlation functions of the plasma, for
a wider variety of geometries~\cite{uno, JM, Tellez-Janco-coulcrit},
and even in the presence of external fields~\cite{Tellez-tcpg}. In
this work, we restrict our analysis to the case $\Gamma=2$.

Assuming $\Gamma=2$, we pretend to find analytical solutions for the
thermodynamics of the two-component plasma in the presence of some
impurities. Results have been obtained for a similar problem, the
one-component plasma with adsorbing impurities~\cite{cuatro,
Cornu-ocp-ad}. For the two-component plasma, some similar problems to
the present one, with adsorbing boundaries, were considered in
Refs.~\cite{siete, Merchan-Tellez-jabon-anillos}.

An impurity in our model can be understood as a particle, different
from the charged particles of the gas. It interacts with the particles
of the gas through a non-electrical potential $U_{\pm}(\r)$. As usual,
for inhomogeneous systems under the action of an external field derived
from a potential $U_{\pm}(\r)$, we can define a position depend
fugacity due to the external potential created by the impurity
$m_{\pm}(\mathbf r)=me^{-\beta U_{\pm}(\mathbf r)}$, where $m$ is the
bulk fugacity. Here we are interested in adsorbing impurities, which
can be modeled with a position depend fugacity $m_{\pm}(\r)=
m\big[1+\alpha_{\pm}\delta(\mathbf r-\mathbf R)\big]$, where
$\alpha_{\pm}$ is the magnitude of the interaction of the impurity
with the positive and the negative particles of the plasma (it will be
called adhesivity), while $\mathbf R$ is the position of the impurity
in the plasma. In order to prevent the collapse of the opposite
charged particles, we will consider that the impurity only attracts
one type of particles of the system. For instance, if the impurity
only attracts the positive particles of the system, $\alpha_+=\alpha$
and $\alpha_-=0$.

This simplified model could be applied to some systems. An example is
a salt solution with positive and negative ions and porous sites,
which can attract some of the ions, with different sizes for each
species. The size of the porous sites can be such that they are small
enough in order to avoid that the larger ions get adsorbed into the
sites, while the smaller ones are allowed to be adsorbed. The same
salt solution may have some ``dusty-points''; the contact interaction
may be generated by a chemical interaction among those points and the
charged particles. An example of a chemical interaction can be
presented in some atoms whose valence layer is not completely filled
and they can receive electrons from an atom with a small
electronegativity. 

The present document is organized as follows. In
section~\ref{sec:method}, we briefly review the general mathematical
treatment which allow us to find the thermodynamic properties and
correlation functions of the two-component plasma in the presence of
any external potential when $\Gamma=2$~\cite{uno}. In
section~\ref{sec:impurities}, we consider the case where there are a
few point impurities in the plasma. We concentrate specifically in the
case of one or two impurities. We compute the density profiles around
the impurities and the two-body correlation functions using the method
of Ref.~\cite{uno}. Also, we extend the treatment presented in
Ref.~\cite{cuatro} for the one-component plasma, to find general
expressions for the grand potential, and the one- and two-body density
functions for the two-component plasma with impurities, in terms of
the same quantities for the unperturbed system. In
section~\ref{sec:lines}, we study the two-component plasma with one or
two lines of adsorbing impurities. We compute the one- and two-density
functions and the grand potential using the method of Ref.~\cite{uno}.
A summary of the main results is presented in
section~\ref{sec:summary}.

\section{Method of solution}
\label{sec:method}

To solve our specific model, we use the method introduced by Cornu and
Jancovici~\cite{uno}. This mathematical treatment is valid for a
globally neutral two-component Coulomb gas with a coupling constant
$\Gamma=2$. Since the model is a two dimensional one, it is convenient
to express the position $\r=(x,y)$ of the particles in complex
coordinates as $z=x+iy$. As explained earlier, at $\Gamma=2$, it is
mandatory to introduce a cutoff distance $a$ between particles to
avoid the collapse between particles of opposite sign. This cutoff can
be understood as a small diameter of the particles. It is shown in
Ref.~\cite{uno} that, in the limit $a\to0$, the grand partition
function $\Xi$ in this scheme is given by
\begin{eqnarray}\label{e1}
\ln \Xi=\textrm{Tr}\left[\ln\left(
\dslash
 +m_+(\mathbf r)\frac{1+\sigma_z}{2}
+m_-(\mathbf r)\frac{1-\sigma_z}{2}\right)\dslash^{-1}\right],
\end{eqnarray}
where $\dslash=\sigma_x\partial_x+\sigma_y\partial_y$ is the two
dimensional Dirac operator. The matrices $\sigma_x,\sigma_y,\sigma_z$
are the Pauli matrices. The position dependent fugacity, defined as
$m_\pm(\mathbf r)=me^{-\beta U_\pm(\mathbf r)}$, takes into account an
external potential $U_{\pm}(\r)$ which can act differently on the
positive or negative particles. The constant rescaled fugacity $m$ is
related to the chemical potential $\mu$ by $m=2\pi e^{\beta \mu}
L/a^2$. It has units of inverse distance, and it is shown in
Ref.~\cite{uno}, that $m^{-1}$ is the screening length.

The calculation of the $k$-body densities is reduced to find the set
of Green functions $G_{s_1s_2}(\mathbf r_1,\mathbf r_2)$, with
$s_1,s_2=\pm1$, of the operator intervening in Eq.~(\ref{e1}). They are
the elements of a $2 \times 2$ matrix which satisfy the differential
equations
\begin{equation}\label{e2}
\left(\begin{array}{cc}
m_+(\mathbf{r}_1) & 2\partial_z\\
2\partial_{\bar z} & m_-(\mathbf r_1)\\
\end{array}\right)\left(\begin{array}{cc}
G_{++}(\mathbf{r}_1,\mathbf{r}_2) & G_{+-}(\mathbf{r}_1,\mathbf{r}_2)\\
G_{-+}(\mathbf{r}_1,\mathbf{r}_2) & G_{--}(\mathbf{r}_1,\mathbf{r}_2)\\
\end{array}\right)=
\delta(\mathbf r_1-\mathbf r_2).
\end{equation}
The density $n_{\pm}(\r)$ and two-body truncated density correlation
function $n^{(2)T}_{\pm}(\r_1,\r_2)$ of the system are given
by~\cite{uno}
\begin{subequations}
\label{eq:correl-Green}
\begin{eqnarray}
\label{e3.1}
n_{s}(\mathbf r)&=&m_s(\mathbf r)G_{ss}(\mathbf r,\mathbf
r)\\
\label{e3.2}
n^{(2)T}_{s_1s_2}(\mathbf r_1,\mathbf r_2)&=&-m_{s_1}(\mathbf r)m_{s_2}(\mathbf r_2)G_{s_1s_2}(\mathbf r_1,\mathbf r_2)G_{s_2s_1}(\mathbf r_2,\mathbf r_1),
\end{eqnarray}
where $T$ means truncated. 
More generally, the truncated $k$-body densities are given
by~\cite{uno}
\begin{eqnarray}
n^{(k)T}_{s_1\ldots s_k}(\r_1,\ldots,\r_k)&=&
(-1)^{k}  m_{s_1}(\r_1)\ldots m_{s_k}(\r_k)
\\
&&\times
\sum_{\text{cycles\ }(i_1 \cdots i_k)}
G_{s_{i_1} s_{i_2}}(\r_{i_1}, \r_{i_2})
\cdots
G_{s_{i_k} s_{i_1}}(\r_{i_k}, \r_{i_1})
\nonumber
\end{eqnarray}
\end{subequations}
The Green functions satisfy the useful relations
\begin{subequations}
\begin{eqnarray}\label{e4.1}
G_{ss}(\mathbf r_1,\mathbf r_2)&=&\overline{G_{ss}(\mathbf r_2,\mathbf
r_1)}\\ 
\label{e4.2}
G_{s-s}(\mathbf r_1,\mathbf
r_2)&=&-\overline{G_{-ss}(\mathbf r_2,\mathbf r_1)}.
\end{eqnarray}
\end{subequations}

The grand potential $\beta\Omega=-\ln\Xi$ can be written as
\begin{eqnarray}\label{e5}
\beta\Omega=-\sum_k\ln\big(1+\lambda_k\big)
\end{eqnarray}
where $\lambda_k$ are the eigenvalues of the system
\begin{eqnarray}\label{e6}
&&\Bigg[\dslash-\frac{1}{\lambda}
\left(\begin{array}{cc}
m_+(\mathbf r) & 0\\
0 & m_-({\mathbf r}) 
\end{array}\right)
\Bigg]\psi(\mathbf r)=0,\;\textrm{with}\nonumber\\
&&\psi(\mathbf r)=
\left(\begin{array}{cc}
g(\mathbf r)\\
f(\mathbf r)
\end{array}\right),
\end{eqnarray}
a two-component spinor. Equation~(\ref{e6}) shows the mentioned equivalence
between the two component plasma at $\Gamma=2$ and a free fermion
gas. The pressure $p$ is given as usual by
\begin{eqnarray}\label{c2.9}
\beta p=\frac{\partial\ln\Xi}{\partial A}=-\frac{1}{A}\beta\Omega,
\end{eqnarray}
in the thermodynamic limit, where $\Omega$ is the grand potential. The
``volume'' of the gas is replaced by its area $A$ because we are
dealing with a two-dimensional model.

The above results are valid in the continuum limit $a\to 0$. However,
in this limit the partition function and other thermodynamical
quantities such as the pressure and the density are divergent. We
shall compute the (divergent) dominant term as a function of the
cutoff $a$. On the other hand, the truncated correlation functions are
finite in the limit $a\to0$.

\section{The plasma with some point adsorbing impurities}
\label{sec:impurities}

In this section, we study the two-component plasma in the presence of
one or two impurities. First, we find the Green functions associated
to this system at $\Gamma=2$. This allows us to find the $k$-body
densities of the plasma. We will distinguish two kind of
impurities. They will be called ``positive'' impurities when they
attract the positive particles of the plasma, and ``negative''
impurities in the opposite case. This is a small abuse of language,
since the impurities do not carry any electrical charge by themselves
alone.

Then we shall extend the general theory presented in
Ref.~\cite{cuatro}, to find general expressions for the thermodynamic
properties of the system with impurities in term of those same
quantities for the unperturbed system. These last results are valid
for any value of the coupling constant $\Gamma$.

\subsection{Density and correlations at $\Gamma=2$}
\label{sec:density-correl-impur}

\subsubsection{A positive impurity}
\label{sec:dens-1impur}

We consider a single impurity in the plasma that only adsorbs the
positive charged particles. In this case the fugacity becomes
$m_+(\mathbf r)=m\big[1+\alpha_1\delta(\mathbf r-\mathbf R_1)\big]$
and $m_-(\mathbf r)=m$. The constants $\alpha_1$ and $\mathbf R_1$
represent the adhesivity of the impurity and its position.

First we compute the Green functions that are needed to find the
density and correlations. To solve the system~(\ref{e2}) for the Green
functions, we assume solutions of the form
$G_{s_1s_2}=G^0_{s_1s_2}+G^1_{s_1s_2}$, where $G^0_{s_1s_2}$ are the
bulk solutions (solutions for an unperturbed plasma with
$m_\pm(\mathbf r)=m$)~\cite{uno}
\begin{subequations}
\label{e7}
\begin{eqnarray}\label{e7.1}
G_{\pm\pm}^0(\mathbf r_1,\mathbf r_2)&=&\frac{m}{2\pi}K_0(m\vert \mathbf r_1 -\mathbf r_2 \vert),\\
G_{-+}^0(\mathbf r_1,\mathbf r_2)&=&\frac{m}{2\pi}\frac{(x_1-x_2)+i(y_1-y_2)}{\vert \mathbf r_1 -\mathbf r_2 \vert}K_1(m\vert \mathbf r_1 -\mathbf r_2 \vert)\nonumber\\
&=&\frac{m}{2\pi}e^{i\theta_{12}}K_1(m\vert \mathbf r_1-\mathbf
r_2\vert)
\,,
\label{e7.2}
\end{eqnarray}
\end{subequations}
where $\theta_{12}$ is the polar angle of the vector $\mathbf
r_1-\mathbf r_2$ and $K_0(x), K_1(x)$ are the modified Bessel
functions of second kind of order 0 and 1. The bulk solutions satisfy
the appropriate boundary condition at $\mathbf r_1=\mathbf r_2$ imposed
by the $\delta(\r_1-\r_2)$ in Eq.~(\ref{e2}). The functions
$G^1_{s_1s_2}$ take into account the contribution of the
impurity. They satisfy the differential equations
\begin{subequations}
\begin{eqnarray}
mG_{++}^1(\r_1,\r_2)+2\partial
_{z_1}G^1_{-+}(\r_1,\r_2)
\nonumber\\
+m\alpha_1\delta(\mathbf r_1 -\mathbf
R_1)\big[G^0_{++}(\r_1,\r_2)+G^1_{++}(\r_1,\r_2)\big]
&=&0
\label{e8.1}
\\ 
2\partial
_{\bar{z}_1}G^1_{++}(\r_1,\r_2)+mG^1_{-+}(\r_1,\r_2)&=&0
\label{e8.2}
\\
mG^1_{+-}(\r_1,\r_2)+
2\partial
_{z_1} G_{--}(\mathbf r_1,\mathbf r_2)
\nonumber\\
+m\alpha_1\delta(\mathbf r_1 -\mathbf
R_1)\big[G^0_{+-}(\r_1,\r_2)+G^1_{+-}(\r_1,\r_2)\big] 
&=&0
\label{e8.3}
\\
2\partial_{\bar{z}_1}G^1_{+-}(\mathbf r_1,\mathbf
r_2)+mG^1_{--}(\mathbf r_1,\mathbf r_2)&=&0.
\label{e8.4}
\end{eqnarray}
\end{subequations}
We will first solve the equations for $G^1_{++}$ and $G^1_{-+}$. We
divide the space into two regions: $r_1<R_1$ and $r_1>R_1$. These
regions will be denoted by the superscripts $(1)$ and $(2)$
respectively. The general solutions for $G_{\pm+}^1$ are
\begin{subequations}
\begin{eqnarray}
\label{e9.1}
G_{++}^1(\r_1,\r_2)&=&\left\{ \begin{array}{cc}
\sum_{l\in\mathbb{Z}} e^{il\theta_1}B^{(1)}_lI_l(mr_1),\;\;\;\;r_1<R_1\\
\sum_{l\in\mathbb{Z}}e^{il\theta_1}A^{(2)}_lK_l(mr_1),\;\;\;\;r_1>R_1
\end{array}\right.\\
G_{-+}^1(\r_1,\r_2)&=&\left\{\begin{array}{cc}
-\sum_{l\in\mathbb{Z}} e^{il\theta_1}B^{(1)}_{l-1}I_l(mr_1),\;\;\;\;r_1<R_1\\
\sum_ {l\in\mathbb{Z}}e^{il\theta_1}A^{(2)}_{l-1}K_l(mr_1),\;\;\;\;r_1>R_1.
\end{array}\right.
\label{e9.2}
\end{eqnarray}
\end{subequations}
with $I_l(x)$ and $K_l(x)$ the modified Bessel functions of first and
second kind of order $l$. From Eq.~(\ref{e8.1}), we notice that
$G_{++}^1$ is continuous, but that $G_{-+}^1$ must be discontinuous at
$\mathbf r_1=\mathbf R_1$, due to the $\delta(\r_1-\mathbf R_1)$
term. Multiplying Eq.~(\ref{e8.1}) by $e^{-il\theta_1}$ and
integrating on $\r_1$ in a small annulus domain, centered at the
origin and containing $\mathbf{R}_1$, gives
\begin{subequations}
\label{e10}
\begin{eqnarray}
G_{++}^1(\mathbf r_1,\mathbf
r_2)\Big\vert_{r_1=R_1^+}&=&G_{++}^1(\mathbf r_1,\mathbf
r_2)\Big\vert_{r_1=R_1^-}\\
\label{e10.1}
\frac{m\alpha_1}{2\pi R_1}\big[G_{++}^0(\mathbf R_1,\mathbf
  r_2)+G_{++}^1(\mathbf R_1,\mathbf r_2)\big]&=&\nonumber\\
&&
\hspace{-4cm} -\Big[G_{-+}^{(1)l+1}(\mathbf r_1,\mathbf
  r_2)\Big\vert_{r_1=R_1^+}-G_{-+}^{(1)l+1}(\mathbf r_1,\mathbf
  r_2)\Big\vert_{r_1=R_1^-}\Big],
\end{eqnarray}
\end{subequations}
where $G_{-+}^{(1)l+1}$ refers to the $l+1$ term of the sum in
Eq.~(\ref{e9.2}). Imposing the boundary conditions~(\ref{e10}), we
find
\begin{equation}
  A_{l}^{(2)}=B^{(1)}_l\frac{I_{l}(mR_1)}{K_{l}(mR_1)}
\end{equation}
and
\begin{equation}\label{e11}
B^{(1)}_l=-\frac{m^3\alpha_1 K_l(mR_1)e^{-il\theta_{R_1}}K_0(m\vert \mathbf R_1-\mathbf r_2\vert)}{4\pi^2\big[1+\frac{m^2\alpha_1}{2\pi}\sum_{n\in\mathbb{Z}}K_n(mR_1)I_n(mR_1)\big]},
\end{equation}
where $\theta_{R_1}$ is the polar angle of the vector $\mathbf
R_1$.

At this point, let us comment a few details about the bulk density,
for an unperturbed system, which will be useful to interpret the
denominator in Eq.~(\ref{e11}). The bulk density is given by
$n_\pm^{0}(\r)=mG_{\pm\pm}^0(\mathbf r,\mathbf r)$. Nevertheless, this
expression diverges, since, for small argument,
$K_0(x)\sim \ln(2/x)-C$, with $C\simeq0.5772$ the Euler
constant~\cite{uno}. This divergence can be avoided by replacing the
point particles by hard spheres of size $a$, which represents the
minimum distance between two charged particles. Then, we compute the
bulk density as $n_{\pm}^{0}= m G_{\pm\pm}^{0}(\r_1,\r_2)$ with
$|\r_1-\r_2|=a\to 0$. Thus,
\begin{equation}
  \label{eq:bulk-dens}
  n_{\pm}^{0}=n_0=\frac{m^2}{2\pi} K_0(m|\r_1-\r_2|)=\frac{m^2}{2\pi} K_0(m a)
  \simeq \frac{m^2}{2\pi}\Big[\ln\frac{2}{ma}-C\Big],
\end{equation}
where we used the small-argument expansion of the Bessel function
$K_0$. Now, let us recall the expansion $K_0(m\vert \mathbf
r_1-\mathbf r_2\vert)=\sum_{n\in\mathbb{Z}}
e^{in(\theta_1-\theta_2)}I_n(mr_<)K_n(mr_>)$ with
$r_{<}=\min(\r_1,\r_2)$ and $r_{>}=\max(\r_1,\r_2)$~\cite{nueve}. With
this expansion, we can obtain another formal (divergent) expression
for the bulk density
\begin{equation}
n_{\pm}^{0}(\mathbf{R}_1)=n_0=\frac{m^2}{2\pi}\sum_{n\in\mathbb{Z}}
K_n(mR_1)I_n(mR_1)\,.
\end{equation}
Since the unperturbed system is homogeneous the bulk density is
constant, $n^{0}_{\pm}(\mathbf{R}_1)=n_0$ does not depend on
$\mathbf{R}_1$. We notice that it is this expression that precisely
appears in the denominator of Eq.~(\ref{e11}). Then,
\begin{equation}\label{e11bis}
B^{(1)}_l=-\frac{m^3\alpha_1 K_l(mR_1)e^{-il\theta_{R_1}}K_0(m\vert
\mathbf R_1-\mathbf
r_2\vert)}{4\pi^2(1+\alpha_1 n_0)}
\,.
\end{equation}
Replacing in Eqs.~(\ref{e9.1}) and~(\ref{e9.2}),
\begin{subequations}\label{e12}
\begin{eqnarray}\label{e12.1}
G_{++}^1(\r_1,\r_2)
=-\frac{m^3\alpha_1}{4\pi^2\big[1+\alpha_1n_0\big]}K_0(m\vert \mathbf R_1-\mathbf r_2\vert)K_0(m\vert \mathbf r_1-\mathbf R_1\vert)\\
\label{e12.2}
G_{-+}^1(\r_1,\r_2)
=-\frac{m^3\alpha_1e^{i\theta_{1R_1}}}{4\pi^2\big[1+\alpha_1n_0\big]}K_0(m\vert
\mathbf r_2-\mathbf R_1\vert)K_1(m\vert \mathbf r_1-\mathbf R_1\vert)
\end{eqnarray}
\end{subequations}
with $\theta_{1R_1}$ the polar angle of the vector $\mathbf r_1-\mathbf R_1$.

In a similar way, the other two Green functions are
\begin{subequations}\label{e13}
\begin{eqnarray}
G_{--}^1(\r_1,\r_2)
&=&\frac{m^3\alpha_1e^{i(\theta_{1R_1}-
\theta_{2R_1})}}{4\pi^2\big[1+\alpha_1n_0\big]}K_1(m\vert\mathbf
r_1-\mathbf R_1\vert)K_1(m\vert\mathbf r_2-\mathbf
R_1\vert)
\nonumber\\
\label{e13.1}
\\
\label{e13.2}
G_{+-}^1(\r_1,\r_2)&=&\frac{m^3\alpha_1
e^{-i\theta_{2R_1}}}{4\pi^2\big[1+\alpha_1n_0\big]}K_0(m\vert \mathbf
r_1-\mathbf R_1\vert)K_1(m\vert \mathbf R_1-\mathbf r_2\vert)
\end{eqnarray}
\end{subequations}
with $\theta_{2R_1}$ the polar angle of the vector $\mathbf
r_2-\mathbf R_1$.  The one-body densities can be calculated by using
Eq.~(\ref{e3.1}), the result is
\begin{subequations}
\label{e14}
\begin{eqnarray}\label{e14.1}
n_+(\mathbf r)&=&\big[1+\alpha_1\delta(\mathbf r-\mathbf R_1)\big]\bigg[n_0-\Big(\frac{m^2}{2\pi}\Big)^2\frac{\alpha_1\big[K_0(m\vert \mathbf R_1-\mathbf r\vert)\big]^2}{[1+\alpha_1n_0\big]}\bigg]\\
\label{e14.2}
n_-(\mathbf r)&=&n_0+\Big(\frac{m^2}{2\pi}\Big)^2\frac{\alpha_1\big[K_1(m\vert \mathbf R_1-\mathbf r\vert)\big]^2}{[1+\alpha_1n_0\big]}.
\end{eqnarray}
\end{subequations}

The number of positive and negative particles of the plasma,
$N_{\pm}=\int n_\pm(\mathbf r)\,d^2\r$, is
\begin{eqnarray}\label{e15}
N_+=N_-=n_0\pi R^2-\frac{m^2\alpha_1}{4\pi\big[1+\alpha_1n_0\big]}+\frac{n_0\alpha_1}{1+\alpha_1n_0},
\end{eqnarray}\\
where we supposed that the plasma is confined in a large disk of
radius $R$. 

\begin{figure}
\includegraphics[width=\GraphicsWidth]{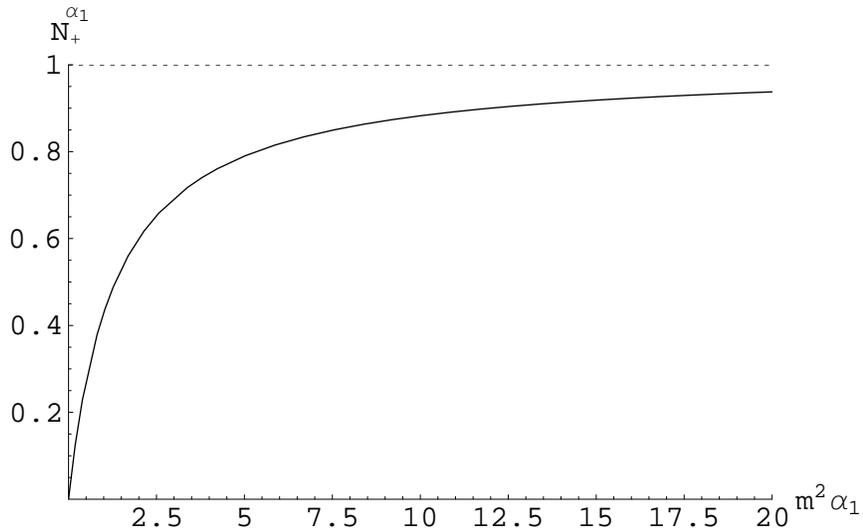}
\caption{
\label{fig:Nplus}
Average number of adsorbed particles as a function of the
adhesivity. The cutoff is $ma=0.01$. The dashed line is the asymptotic
value $N_{+}^{\infty}=1$ for $\alpha_1\rightarrow\infty$.}
\end{figure}

The difference $N_+-N_-=0$. This shows that the global neutrality of
the system is not changed, only the charge distribution. The amount of
positive charge bound to the impurity, $N_{+}^{\alpha_1}$, is given by
the term which multiply the delta distribution in Eq.~(\ref{e14.1}),
evaluated at $\mathbf{R}_1$. The result is
\begin{eqnarray}\label{e15a}
N^{\alpha_1}_+=\frac{\alpha_1n_0}{1+\alpha_1n_0},
\end{eqnarray}
which can be interpreted as a mean occupation number or probability
that the adsorbing point is occupied by a positive
charge. Figure~\ref{fig:Nplus} shows $N_{+}^{\alpha_1}$ as a function
of $\alpha_1$. As the adhesivity $\alpha_1$ is increased, the average
number of adsorbed positive particles $N_{+}^{\alpha_1}$ increases, as
expected. In the limit $\alpha_1\to\infty$, the maximum value of
adsorbed particles is obtained $N_{+}^{\infty}=1$. Notice that in
average there cannot be more than one adsorbed particle per impurity
site. Once a particle is adsorbed, any other particle of the same sign
feels a strong electrostatic repulsion that prevent it from
approaching the adsorbtion site.

\begin{figure}
\begin{center}
\includegraphics[width=\GraphicsWidth]{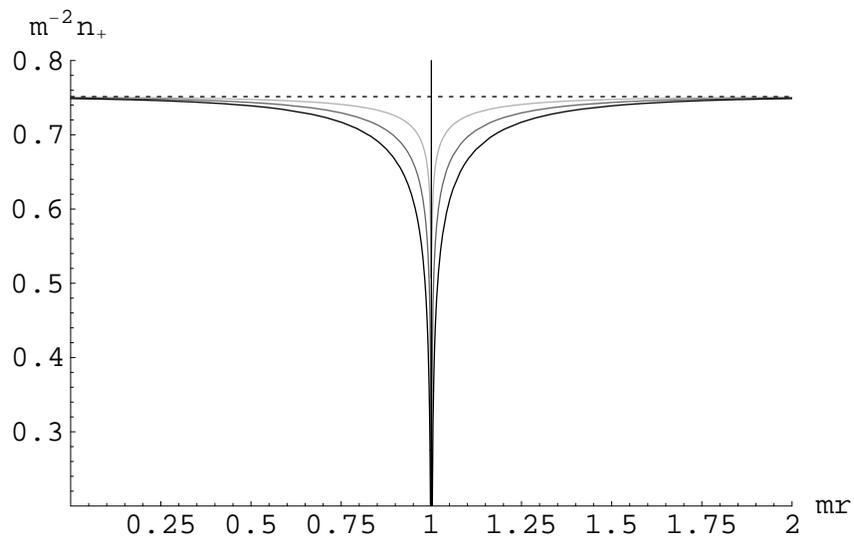}
\caption{
\label{fig:plus1}
Density of positive particles as a function of the distance for
$m^2\alpha_1=1,0.5$ and $0.2$ from the darkest to the lightest. The
dashed line represents the bulk density. The vertical line at $mr=1$
represents the Dirac distribution related to the impurity position
which is located at $m\mathbf R_1=\hat x$. The angular difference is
$\theta-\theta_{R_1}=0$. The cutoff is $ma=0.01$.}
\end{center}
\end{figure}

\begin{figure}
\begin{center}
\includegraphics[width=\GraphicsWidth]{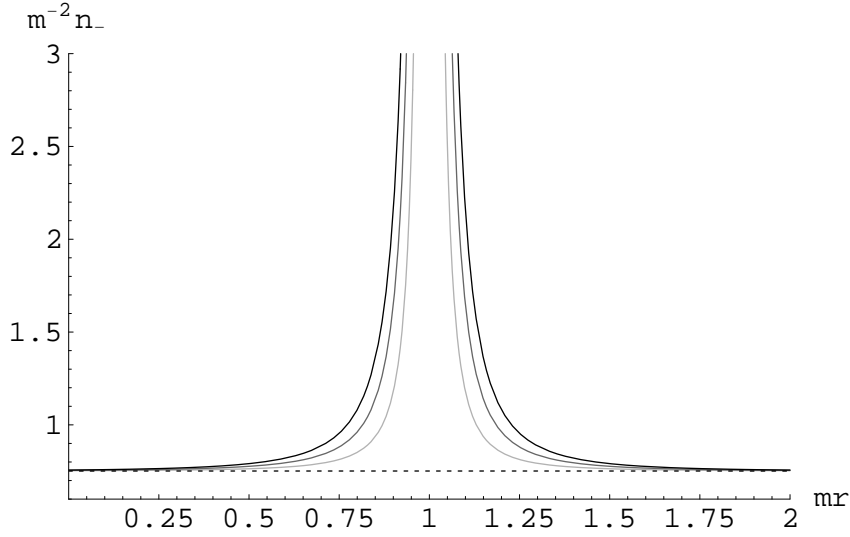}
\caption{\label{fig:minus1} Density of negative particles as a
function of the distance for $m^2\alpha_1=1,0.5$ and $0.2$ from the
darkest to the lightest. The dashed line represents the bulk
density. The angular difference is $\theta-\theta_{R_1}=0$. The cutoff
is $ma=0.01$.}
\end{center}
\end{figure}

\begin{figure}
\begin{center}
\includegraphics[width=\GraphicsWidth]{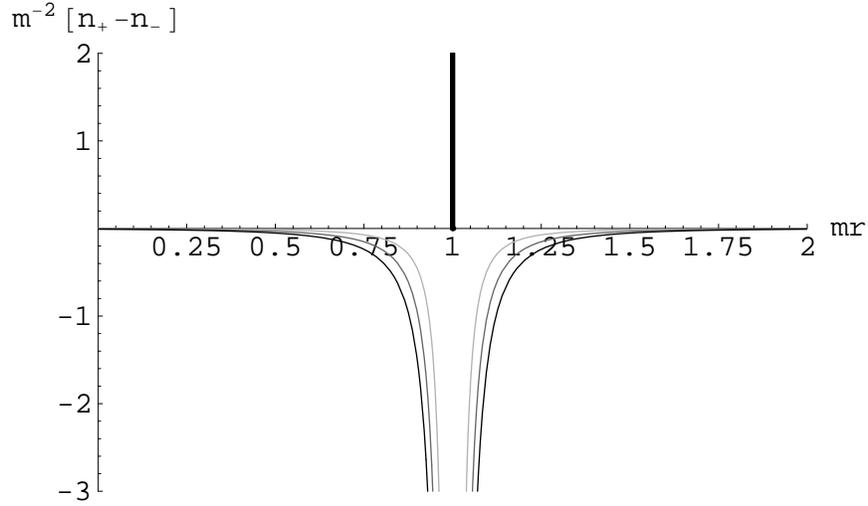}
\caption{\label{fig:charge-density1} Charge density profile as a
function of the distance for $m^2\alpha_1=1,0.5$ and $0.2$ from the
darkest to the lightest. The vertical line at $mr=1$ represents the
Dirac distribution related to the impurity position which is located
at $m\mathbf R_1=\hat x$. The angular difference is
$\theta-\theta_{R_1}=0$. The cutoff is $ma=0.01$.}
\end{center}
\end{figure}

This localized positive adsorbed charge is the responsible of the
change in the charge distribution. Since the bound charge is positive,
it repels the positive particles and attracts the negative ones. As a
consequence, the density of the negative particles around the impurity
increases, and the density of the positive particles around the
impurity decreases. As it is expected, the larger the adhesivity
$\alpha_1$ is, the larger this effect is. This behavior can be seen in
figures~\ref{fig:plus1}, \ref{fig:minus1}
and~\ref{fig:charge-density1} which show the density profiles of the
positive and negative particles, and the charge density profile.

\begin{figure}
  \begin{center}
    \includegraphics[width=\GraphicsWidth]{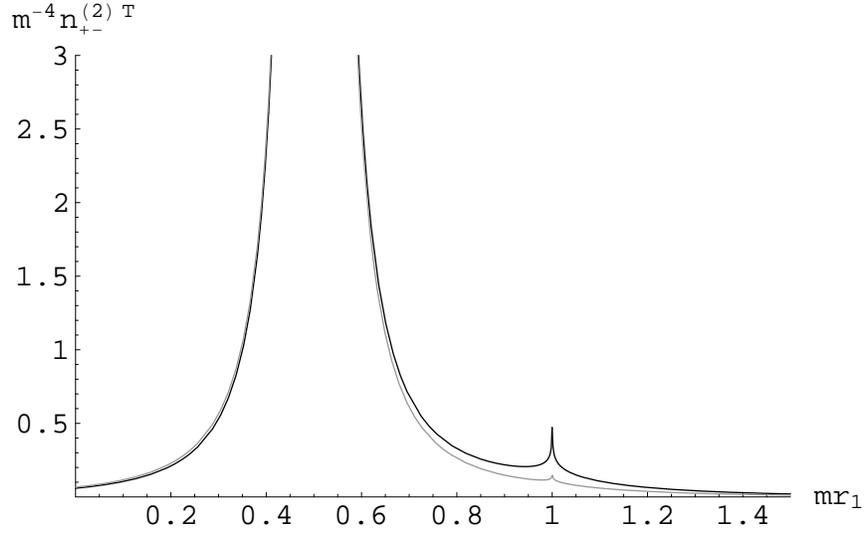}
    \caption{
      \label{fig:npm-1impur}
      $n^{(2)T}_{+-}(\r_1,\r_2)$ as a function of $r_1$ for $m^2\alpha_1=10$
      and $0.5$ from the darkest to lightest. We assumed $mr_2=0.5$ and
      $mR_1=1$. The angular differences are
      $\theta_1-\theta_2=\theta_1-\theta_{R_1}=0$: we are looking through
      a line which passes over $\mathbf r_2$ and $\mathbf
      R_1$. The delta function at $\mathbf r_1=\mathbf R_1$ is not showed.}
  \end{center}
\end{figure}

\begin{figure}
\begin{center}
\includegraphics[width=\GraphicsWidth]{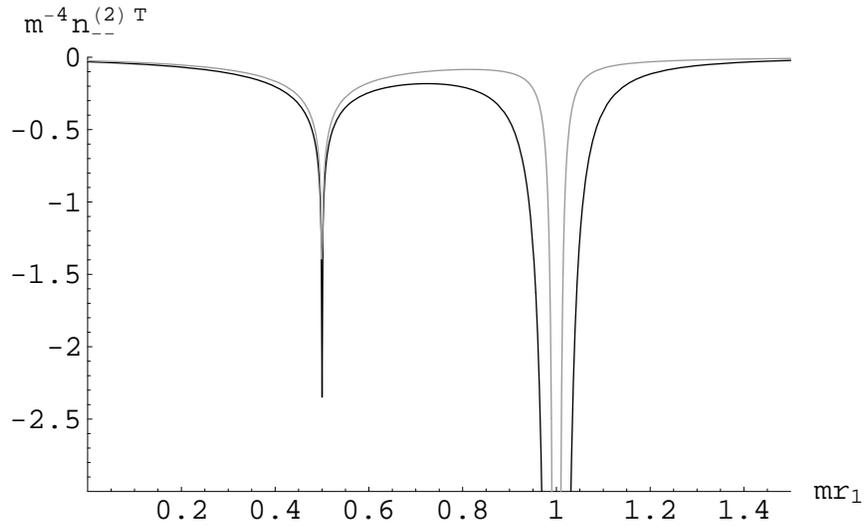}
\caption{\label{fig:nmm-1impur}$n^{(2)T}_{--}(\r_1,\r_2)$ as a function of the
distance. We assume the same parameter values as in
figure~\ref{fig:npm-1impur}.}
\end{center}
\end{figure}

\begin{figure}
\begin{center}
\includegraphics[width=\GraphicsWidth]{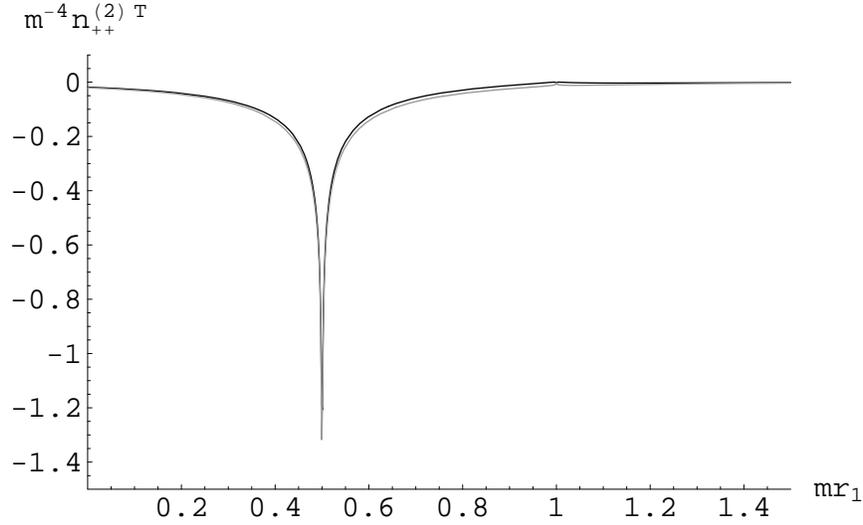}
\caption{
\label{fig:npp-1impur}
$n^{(2)T}_{++}(\r_1,\r_2)$ as a function of the distance. We assume
the same parameter values as in figure~\ref{fig:npm-1impur}. The delta function
is not plotted.}
\end{center}
\end{figure}

If the impurity attracts the negative particles, instead of the
positive ones, we obtain the same results by interchanging ($+
\longleftrightarrow -$). 

As seen in figures~\ref{fig:plus1}, \ref{fig:minus1}
and~\ref{fig:charge-density1}, a charge polarization cloud is formed
around the impurity. It is interesting to evaluate the electric
potential created by this charge distribution. To compute the electric
potential, we use the Poisson equation $\nabla^2\phi(\mathbf r)=-2\pi
e[n_+(\mathbf r)-n_-(\mathbf r)]$, where $n_{\pm}(\mathbf r)$ are
given in Eqs.~(\ref{e14}). In the thermodynamic limit the plasma is
large enough and we can assume an angular symmetry around the impurity
(locating it at the origin). Thus, Poisson equation reduces to
\begin{eqnarray}\label{e16}
\frac{1}{r}\frac{d}{dr}\Big[r\frac{d}{dr}\phi(r)\Big]=-2\pi e\big[n_+(r)-n_-(r)\big].
\end{eqnarray}
Integrating, we find the electric potential 
\begin{equation}\label{e17}
\phi(r)=\frac{em^2\alpha_1}{2\pi\big[1+\alpha_1n_0\big]}\big[K_0(mr)\big]^2
=\frac{e m^2 N_{+}^{\alpha_1}}{2n_0\pi}
\big[K_0(mr)\big]^2.
\end{equation}
From Eq.~(\ref{e17}) we can see that close to the impurity, the
potential behaves as $\big[\ln(mr)\big]^2$, which is stronger than the
bare Coulomb potential $-\ln (r/L)$. This is due to the strong
electrostatic coupling considered here~\cite{GT-small-r},
for smaller coupling, one would expect that the electric potential
close to the impurity behaves as the bare Coulomb potential. Far away
from it, the interaction decays exponentially due to the screening
effect.

Replacing the results for the Green functions, Eqs.~(\ref{e12})
and~(\ref{e13}), into Eq.~(\ref{e3.2}) allows us to obtain the
two-body density correlation functions. Figures~\ref{fig:npm-1impur},
\ref{fig:nmm-1impur}, and \ref{fig:npp-1impur} show plots of the
two-body densities, for different values of the adhesivity.

\subsubsection{Two positive impurities}
\label{sec:dens-2impur-pp}

Now, we consider two impurities located at $\mathbf{R}_1$ and
$\mathbf{R}_2$ with adhesivities $\alpha_1$ and $\alpha_2$
respectively, both attracting positive particles. The fugacities are
now given by $m_+(\mathbf r)=m\big[1+\alpha_1\delta(\mathbf r-\mathbf
R_1)+\alpha_2\delta(\mathbf r-\mathbf R_2)\big]$ and $m_-(\mathbf
r)=m$.

To solve the system of equations~(\ref{e2}) for the Green functions,
we assume they are of the form
$G_{s_1s_2}=G^0_{s_1s_2}+G^1_{s_1s_2}+G^2_{s_1s_2}$ where
$G^1_{s_1s_2}$ are the solutions previously found for one impurity
located at $\mathbf{R}_1$. Without lost of generality we choose $R_2>
R_1$. The space is now divided into three regions. The first region is
$r_1<R_1$, the second region is $R_1<r_1<R_2$ and the third one is
$R_2<r_1$. The differential equations satisfied by $G_{\pm+}^{2}$ are
\begin{subequations}
\begin{eqnarray}\label{e18.1}
m\big[1+\alpha_1 \delta(\mathbf r_1 -\mathbf
  R_1)+\alpha_2\delta(\mathbf r_1 -\mathbf
  R_2)\big]G_{++}^2(\r_1,\r_2)
&&\nonumber\\
+m\alpha_2\delta(\mathbf r_1 -\mathbf
R_2)\big[G_{++}^0(\r_1,\r_2)+G_{++}^1(\r_1,\r_2)\big]
\nonumber\\
+2\partial_{z_1}G_{-+}^2(\r_1,\r_2)
&=&0
\\
2\partial_{\bar{z}_1}G_{++}^2(\r_1,\r_2)+mG_{-+}^2(\r_1,\r_2)&=&0.
\label{e18.2}
\end{eqnarray}  
\end{subequations}
with general solutions of the form
\begin{eqnarray}\label{e19}
G_{++}^2(\r_1,\r_2)&=&\left\{ 
\begin{array}{ll}
\sum_{l\in\mathbb{Z}}
e^{il\theta_1}B_l^{(1)}I_l(mr_1) & r_1<R_1\\
\sum_{l\in\mathbb{Z}}
e^{il\theta_1}
\left[
B_l^{(2)}I_l(mr_1)
+A_l^{(2)}K_l(mr_1)\right]& R_1<r_1<R_2\\
\sum_{l\in\mathbb{Z}}
e^{il\theta_1}A_l^{(3)}K_l(mr_1)& R_2<r_1
\end{array}\right.\nonumber\\
G_{-+}^2(\r_1,\r_2)&=&\left\{\begin{array}{ll}
-\sum_{l\in\mathbb{Z}}
e^{il\theta_1}B_{l-1}^{(1)}I_l(mr_1)& r_1<R_2\\
\sum_{l\in\mathbb{Z}}
e^{il\theta_1}\left[
-B_{l-1}^{(2)}I_l(mr_1)+
A_{l-1}^{(2)}K_l(mr_1)\right]& R_1<r_1<R_2\\
\sum_{l\in\mathbb{Z}}
e^{il\theta_1}A_{l-1}^{(3)}K_l(mr_1)& R_2<r_1\,
\end{array}\right.\nonumber\\
\end{eqnarray}
The delta functions in Eq.~(\ref{e18.1}) impose that $G^2_{++}$ is
continuous at $\mathbf r_1=\mathbf R_1$ and $\mathbf r_1=\mathbf R_2$
but $G^2_{-+}$ is discontinuous. Imposing these four boundary
conditions, we find after some algebra
\begin{eqnarray}\label{e20}
A_l^{(2)}&=&-\frac{m^4\alpha_1\alpha_2K_0(m\vert\mathbf R_1-\mathbf R_2\vert)}{4\pi^2\eta}I_l(mR_1)e^{-il\theta_{R_1}}D_1,\nonumber\\
A_l^{(3)}&=&\frac{m^2\alpha_2}{2\pi\eta}I_l(mR_2)e^{-il\theta_{R_2}}\big[1+\alpha_1n_0\big]D_1+A_l^{(2)},\nonumber\\
B_l^{(2)}&=&\frac{m^2\alpha_2}{2\pi\eta}K_l(mR_2)e^{-il\theta_{R_2}}\big[1+\alpha_1n_0\big]D_1,\nonumber\\
B_l^{(1)}&=&B_l^{(2)}-\frac{m^4\alpha_1\alpha_2K_0(m\vert\mathbf R_1-\mathbf R_2\vert)}{4\pi^2\eta}K_l(mR_1)e^{-il\theta_{R_1}}D_1.
\end{eqnarray}
with
\begin{eqnarray}
D_1&=&-\frac{m}{2\pi}K_0(m\vert \mathbf r_2-\mathbf
R_2\vert)+\frac{m^3\alpha_1 K_0(m\vert \mathbf R_1- \mathbf
  r_2\vert)K_0(m\vert \mathbf R_1-\mathbf
  R_2\vert)}{4\pi^2\big[1+\alpha_1n_0\big]},
\nonumber\\
\label{e21}\\
\eta&=&1+\alpha_1n_0+\alpha_2n_0+\alpha_1\alpha_2n_0^2-\frac{m^4\alpha_1\alpha_2}{4\pi^2}\big[K_0(m\vert\mathbf
  R_1-\mathbf R_2\vert)\big]^2.
\label{eq:eta}
\end{eqnarray}
The functions $G_{\pm+}^2$ are therefore
\begin{subequations}
\label{e22}
\begin{eqnarray}
G_{++}^2(\r_1,\r_2)&=&\frac{m^2\alpha_2}{2\pi\eta}D_1\Bigg[\big[1+\alpha_1n_0\big]K_0(m\vert\mathbf r_1-\mathbf R_2\vert)\nonumber\\
&-&{\frac{m^2\alpha_1}{2\pi}}K_0(m\vert\mathbf r_1-\mathbf R_1\vert)K_0(m\vert\mathbf R_1-\mathbf R_2\vert)\Bigg]\\
G_{-+}^2(\r_1,\r_2)&=&\frac{m^2\alpha_2}{2\pi\eta}D_1\Bigg[\big[1+\alpha_1n_0\big]K_1(m\vert\mathbf r_1-\mathbf R_2\vert)e^{i\theta_{1R_2}}\nonumber\\
&-&{\frac{m^2\alpha_1}{2\pi}}K_1(m\vert\mathbf r_1-\mathbf R_1\vert)e^{i\theta_{1R_1}}K_0(m\vert\mathbf R_1-\mathbf R_2\vert)\Bigg].
\end{eqnarray}
\end{subequations}
Following some similar steps, we find the other two Green functions
\begin{subequations}
\label{e23}
\begin{eqnarray}
G_{--}^2(\r_1,\r_2)&=&-\frac{m^2\alpha_2}{2\pi\eta}D_2\Bigg[\big[1+\alpha_1n_0\big]K_1(m\vert\mathbf r_1-\mathbf R_2\vert)e^{i\theta_{1R_2}}\nonumber\\
&-&\frac{m^2\alpha_1}{2\pi}K_1(m\vert\mathbf r_1-\mathbf R_1\vert)e^{i\theta_{1R_1}}K_0(m\vert\mathbf R_1-\mathbf R_2\vert)\Bigg],\nonumber\\
G_{+-}^2(\r_1,\r_2)&=&-\frac{m^2\alpha_2}{2\pi\eta}D_2\Bigg[\big[1+\alpha_1n_0\big]K_0(m\vert\mathbf r_1-\mathbf R_2\vert)\nonumber\\
&-&\frac{m^2\alpha_1}{2\pi}K_0(m\vert\mathbf r_1-\mathbf R_1\vert)K_0(m\vert\mathbf R_1-\mathbf R_2\vert)\Bigg],
\end{eqnarray}
\end{subequations}
with
\begin{eqnarray}\label{e24}
D_2&=&-\frac{m}{2\pi}K_1(m\vert \mathbf R_2-\mathbf r_2\vert)e^{-i\theta_{2R_2}}\nonumber\\
&+&\frac{m^3\alpha_1 K_1(m\vert \mathbf R_1-\mathbf r_2\vert)K_0(m\vert\mathbf R_2-\mathbf R_1\vert)e^{-i\theta_{2R_1}}}{4\pi^2\big[1+\alpha_1n_0\big]}.
\end{eqnarray}

\begin{figure}
\includegraphics[width=\GraphicsWidth]{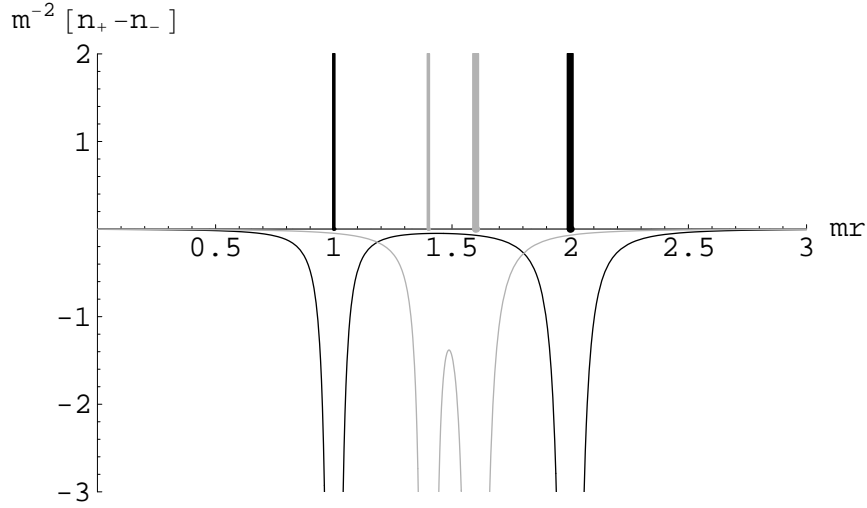}
\caption{
\label{fig:charge-2impur-plus-plus}
Charge density profile for two ``positive'' impurities. The
adhesivities are $m^2\alpha_1=0.2$ and $m^2\alpha_2=0.5$. The
positions of the impurities are on the $x$ axis at $m R_1=1$,
$mR_1=1.4$ and $m R_2=2$, $mR_2=1.6$, respectively from the darkest to
the lightest. All the angular differences are zero. The cutoff is
$ma=0.01$.}
\end{figure}

Replacing these Green functions in Eq.~(\ref{e3.1}) gives the density
profiles. Figure~\ref{fig:charge-2impur-plus-plus} shows the charge
density profile. It can be seen that the charge density decreases
around the adsorbing points like it was found in the problem with one
impurity. The density depends on the adhesivity as well as of the
distance between the impurities. An important effect arises when the
adsorbing points are close enough which indicates the existence of an
indirect interaction between the impurities (see
section~\ref{sec:two-impurities-interaction}).

\subsubsection{One positive and one negative impurity}
\label{sec:dens-2impur-pm}

Now we consider the case when the second impurity only adsorbs
the negative particles. The fugacities for this configuration are
$m_+(\mathbf r)=m\big[1+\alpha_1\delta(\mathbf r-\mathbf R_1)\big]$
and $m_-(\mathbf r)=m\big[1+\alpha_2\delta(\mathbf r-\mathbf
R_2)\big]$. Assuming a similar solution as in Eq.~(\ref{e19}), and
after establishing the appropriate boundary conditions for this case,
we solve the linear system~(\ref{e2}), to find
\begin{eqnarray}\label{e25}
G_{++}(\r_1,\r_2)&=&
G_{++}^0(\r_1,\r_2)
+G_{++}^1(\r_1,\r_2)
\nonumber\\
&+&
\frac{m^2\alpha_2}{2\pi\bar\eta}D_3\Bigg[\big[1+\alpha_1n_0\big]
K_1(m\vert\mathbf{r_1}-\mathbf R_2\vert)e^{-i\theta_{1R_2}}
\nonumber\\
&-&\frac{m^2\alpha_1}{2\pi}K_0(m\vert\mathbf r_1-\mathbf
  R_1\vert)e^{-i\theta_{R_1R_2}}K_1(m\vert\mathbf R_1-\mathbf
  R_2\vert)\Bigg],
\\
G_{-+}(\r_1,\r_2)&=&G_{-+}^0(\r_1,\r_2)+G_{-+}^1
(\r_1,\r_2)
\\
&-&\frac{m^2\alpha_2}{2\pi\bar\eta}D_3\Bigg[\big[1+\alpha_1n_0\big]e^{i\theta_{R_2}}K_0(m\vert\mathbf r_1-\mathbf R_2\vert)e^{-i\theta_{1R_2}}\nonumber\\
&+&\frac{m^2\alpha_1}{2\pi}K_1(m\vert\mathbf r_1-\mathbf R_1\vert)
e^{i(\theta_{1R_1}-\theta_{R_1R_2})}
K_1(m\vert\mathbf R_1-\mathbf R_2\vert)\Bigg],
\nonumber
\end{eqnarray}
with
\begin{eqnarray}\label{e26}
D_3&=&\frac{m^2}{2\pi}e^{i\theta_{2R_2}}K_1(m\vert\mathbf R_2-\mathbf r_2\vert)\nonumber\\
&-&\frac{m^3\alpha_1 K_0(m\vert \mathbf R_1-\mathbf r_2\vert)e^{i\theta_{R_1R_2}}K_1(m\vert\mathbf R_2-\mathbf R_1\vert)}{4\pi^2\big[1+\alpha_1n_0\big]},\\
\bar\eta&=&1+\alpha_1n_0+\alpha_2n_0+\alpha_1\alpha_2n_0^2+\frac{m^4\alpha_1\alpha_2}{4\pi^2}\big[K_1(m\vert\mathbf
  R_1-\mathbf R_2\vert)\big]^2.
\label{eq:etabar}
\end{eqnarray}
and $\theta_{R_1R_2}$ the polar angle of the vector $\mathbf
R_1-\mathbf R_2$.

In order to find $G_{--}$, we note that if we interchange $\mathbf
R_1$ and $\mathbf R_2$ and their adhesivities in the problem for
$G_{++}$, we would have a negative impurity at $\mathbf r_1=\mathbf
R_1$, and a positive one at $\mathbf r_1=\mathbf R_2$. Due to this
symmetry argument, the function $G_{--}$ for one positive and one
negative impurity at $\mathbf r_1=\mathbf R_1$ and $\mathbf
r_1=\mathbf R_2$, is the function $G_{++}$ with $\mathbf R_1$ and
$\mathbf R_2$ and their adhesivities interchanged. Then,
\begin{eqnarray}\label{e27}
G_{--}(\mathbf r_1,\mathbf r_2;\mathbf R_1,\alpha_1,\mathbf R_2,\alpha_2)=G_{++}(\mathbf r_1,\mathbf r_2;\mathbf R_2,\alpha_2,\mathbf R_1,\alpha_1).
\end{eqnarray}
with $G_{++}$ given in Eq.~(\ref{e25}). 

\begin{figure}
\includegraphics[width=\GraphicsWidth]{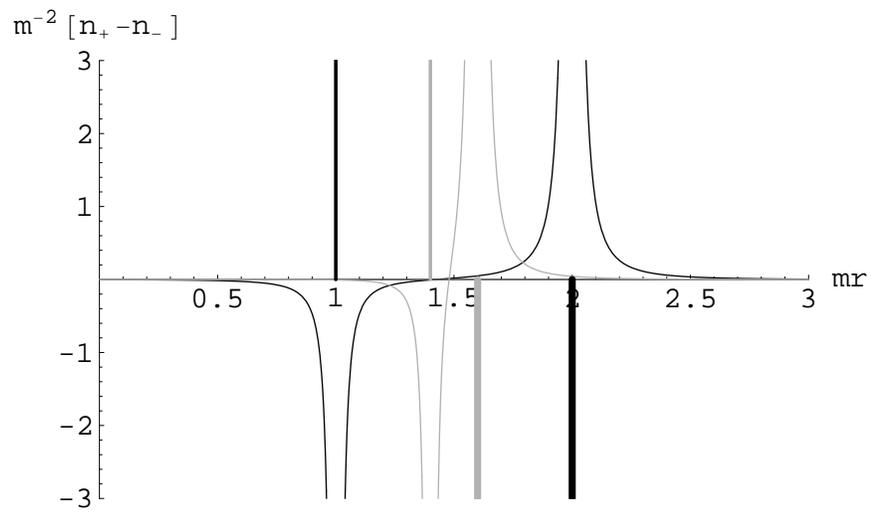}
\caption{
\label{fig:charge-2impur-plus-minus}
Charge density profile for one ``positive'' impurity and one
``negative'' impurity. The adhesivities are $m^2\alpha_1=0.2$
(positive impurity) and $m^2\alpha_2=0.5$ (negative impurity). The
positions of the impurities are on the $x$ axis at $m R_1=1$,
$mR_1=1.4$ (positive impurity) and $m R_2=2$, $mR_2=1.6$ (negative
impurity), respectively from the darkest to the lightest. All the
angular differences are zero. The cutoff is $ma=0.01$.}
\end{figure}

Figure~\ref{fig:charge-2impur-plus-minus} shows the charge density
profile for this situation.

\subsubsection{General structure of the Green functions for two
  impurities}
\label{sec:dens-2impur-gen}

The results of the two previous sections can be put in more general
and compact form. Let $S_1=\pm 1$ be the sign of the particles the
impurity at $\mathbf{R}_1$ attracts and $S_2=\pm 1$ the sign of the
particles the second impurity at $\mathbf{R}_2$ attracts. Let us
denote by $\hat G_{s_1 s_2}=G^0_{s_1 s_2}+G^1_{s_1 s_2}$, the Green
functions for the case of only one impurity located at $\mathbf{R}_1$
(Eqs.~(\ref{e12}) and~(\ref{e13})). Notice that in general
Eqs.~(\ref{e12}) and~(\ref{e13}) can be written as
\begin{equation}
  \label{eq:G-1impur-gen}
  \hat{G}_{s_1 s_2}(\r_1,\r_2)=
  G^{0}_{s_1 s_2}(\r_1,\r_2)
  -\frac{m\alpha_1}{1+\alpha_1 n_{S_1}^{0}(\R_1)}
  G^{0}_{s_1 S_1}(\r_1,\R_1)  G^{0}_{S_1 s_2}(\R_1,\r_2)
\end{equation}
in terms of the Green functions $G^{0}_{s_1 s_2}$ and the density
$n_{s}^{0}$ of the unperturbed system.

The solutions for the Green functions found in the two previous
sections for the case of two impurities can be written, in term of the
Green function for one impurity, as
\begin{equation}
  G_{s_1 s_2}(\r_1,\r_2)=
  \hat{G}_{s_1 s_2}(\r_1,\r_2)
  -\frac{m\alpha_2}{\eta}
  \hat{G}_{s_1 S_2}(\r_1,\R_2)
  \hat{G}_{S_2 s_2}(\R_2,\r_2)
  \label{eq:G-2impur-1impur}
\end{equation}
where $\eta$ was defined in Eqs.~(\ref{eq:eta}) and~(\ref{eq:etabar})
(denoted $\bar\eta$ in this last equation). In general
\begin{equation}
  \label{eq:etagen}
  \eta=1
  +\alpha_1 n_{S_1}^{0}(\R_1)
  +\alpha_2 n_{S_2}^{0}(\R_2)
  +\alpha_1 \alpha_2 n^{0,(2)}_{S_1 S_2}(\R_1,\R_2)
\end{equation}
where $n^{0,(2)}_{S_1 S_2}(\R_1,\R_2)$ is the (non-truncated) pair
correlation function for the unperturbed system. Notice the similarity
in the structure of the Green functions in
Eqs.~(\ref{eq:G-1impur-gen}) and~(\ref{eq:G-2impur-1impur}).

Replacing Eq.~(\ref{eq:G-1impur-gen}) into
Eq.~(\ref{eq:G-2impur-1impur}) gives the expression
\begin{eqnarray}
  G_{s_1 s_2}(\r_1,\r_2)&=& G^0_{s_1 s_2}(\r_1,\r_2)
  \nonumber\\
  &-&\frac{1}{\eta}
  \Bigg[
    m\alpha_1 [1+\alpha_2 n_{S_2}^{0}(\R_2)]
    G_{s_1 S_1}^{0}(\r_1,\R_1) G^{0}_{S_1 s_2}(\R_1,\r_2)
    \nonumber\\
    &&
    + m\alpha_2 [1+\alpha_1 n_{S_1}^{0}(\R_1)]
    G_{s_1 S_2}^{0}(\r_1,\R_2) G^{0}_{S_2 s_2}(\R_2,\r_2)
    \nonumber\\
    &&
    - m^2 \alpha_1 \alpha_2
    \big[ 
      G_{s_1 S_1}^{0} (\r_1,\R_1)
      G_{S_1 S_2}^{0} (\R_1,\R_2)
      G_{S_2 s_2}^{0} (\R_2,\r_2)
      \nonumber\\
      &&
      \hspace{1cm}+
      G_{s_1 S_2}^{0} (\r_1,\R_2)
      G_{S_2 S_1}^{0} (\R_2,\R_1)
      G_{S_1 s_2}^{0} (\R_1,\r_2)
      \big]
    \Bigg]
  \nonumber\\
  \label{eq:Green-gen-2impur}
\end{eqnarray}
in which the exchange symmetry between $(\R_1,S_1)$ and $(\R_2,S_2)$
is manifest.

\subsection{General results}

The grand potential of the system with impurities can be found by
solving the eigenvalue problem~(\ref{e6}) for each particular
situation. We will proceed this way to compute the grand potential for
the Coulomb gas in the presence of a continuous adsorbing line in
subsections~\ref{sec:gp-1line} and~\ref{sec:gp-2lines}. Although this
method can also be used to compute the grand potential with point
impurities~\cite{tesis-Ferrero}, for this latter case, we shall
proceed differently, using the general theory developed in
Ref.~\cite{cuatro} for a one-component fluid with adsorbing sites.

In the following subsections, we obtain expressions for the grand
potential and the one- and two-body densities of the plasma with
impurities in terms of those same quantities for the plasma without
impurities (the unperturbed system). These results are very general:
they are valid for any value of the coupling $\Gamma$, they are even
valid for any other type of interaction potential between the
particles of the system, not only for the Coulomb system considered
here. In the present case, since at $\Gamma=2$ the grand potential
and density functions for the unperturbed system are known, one can
obtain explicit expressions for the thermodynamics and correlations of
the plasma with impurities. 

\subsubsection{Grand partition function with an arbitrary number of impurities}

Suppose there are $M_+$ impurities attracting positive particles,
located at $\R_1^+, \ldots, \R_{M_{+}}^+$, with adhesivities
$\alpha_{1}^{+}, \ldots, \alpha_{M_{+}}^{+}$, and $M_-$ impurities
attracting negative particles, located at $\R_1^-, \ldots,
\R_{M_{-}}^-$, with adhesivities $\alpha_{1}^{-}, \ldots,
\alpha_{M_{-}}^{-}$. The general calculations done in
Ref.~\cite{cuatro} for a one-component plasma, can directly be adapted
to the present two-component case, to find that the grand partition
function $\Xi$ can be written as
\begin{eqnarray}
  \Xi&=&\Xi_0
  \sum_{s_+=0}^{\infty}   \sum_{s_-=0}^{\infty}
  \sum_{k_{1}^{+},\ldots,k_{s_{+}}^{+}=0}^{M_{+}}
  \sum_{k_{1}^{-},\ldots,k_{s_{-}}^{-}=0}^{M_{-}}
  \frac{
  \prod_{n=1}^{s_{+}} \alpha_{k_{n}^{+}}
  \prod_{n=1}^{s_{-}} \alpha_{k_{n}^{-}}
  }{s_{+}! s_{-}!}
  \nonumber\\
  &&
  \hspace{1cm}
  n^{0,(s_{+}+s_{-})}_{\{s_+\}\{s_-\}}
  (\R_{k_1^+}^{+},\ldots,\R_{k_{s_{+}}^+}^{+},
  \R_{k_1^-}^{-},\ldots,\R_{k_{s_{-}}^-}^{-})
  \label{eq:gp-general}
\end{eqnarray}
where $\Xi_0$ is the grand partition function of the unperturbed system
and $n^{0,(s_{+}+s_{-})}_{\{s_+\}\{s_-\}}$ the $(s_{+}+s_{-})$-body
density for $s_{+}$ positive particles and $s_{-}$ negative particles
of the unperturbed system. Notice that, since the correlation function
vanishes if two of its arguments are equal, the above expression is not
an infinite sum, it contains at most terms involving the
$(M_{+}+M_{-})$-body correlation function and lower degree
correlations.

From expression~(\ref{eq:gp-general}) we can obtain the density
and $k$-body correlation functions performing successive functional
derivations of the grand partition function with respect to the
fugacity.

\subsubsection{One impurity}
\label{sec:gp-1impur}

In the case of a single impurity located at $\R_1$, with adhesivity
$\alpha_1$ and attracting particles of sign $S_1$,
Eq.~(\ref{eq:gp-general}) simply reduces to
\begin{equation}
  \label{eq:Xi-1impur}
  \Xi= \big[1+\alpha_1 n_{S_1}(\R_1)\big] \Xi_0
  \,.
\end{equation}
Then, the grand potential $\Omega=-k_B T \ln \Xi$ can be written as
$\Omega=\Omega_0+\Omega_{\exc}(\alpha_1)$ with $\Omega_0$ the grand potential
of the unperturbed system and an excess grand potential
\begin{equation}
  \beta \Omega_{\exc}(\alpha_1)=-\ln\big[1+\alpha_1 n_{S_1}(\R_1)\big]
  \,.
\end{equation}

As explained in Ref.~\cite{cuatro}, the adhesivity $\alpha$ can be
interpreted also as a fugacity for the adsorbed particles. Thus, one
can compute the average number of adsorbed particles from the relation
\begin{equation}
  N_{S_1}^{\alpha_1}=-\alpha_1\frac{\partial 
    \beta\Omega_{\exc}(\alpha_1)}{\partial \alpha_1}=
  \frac{\alpha_1 n_{S_1}^{0}(\R_1)}{1+\alpha_1 n^{0}_{S_1}(\R_1)}
  \,.
\end{equation}
We recover the result~(\ref{e15a}) obtained from a direct calculation
at $\Gamma=2$.

From Eq.~(\ref{eq:Xi-1impur}) we can also rederive the results from
the previous section for the density and correlation functions, by
performing functional derivatives of $\Xi$ with respect to the
fugacity. Remembering that
\begin{equation}
  m_s(\r)\left.\frac{\delta \ln \Xi_0}{\delta m_s(\r)}
  \right|_{m(\r)=m}=n^{0}_{s}(\r)
\end{equation}
and
\begin{equation}
  m_{s_2}(\r_2)\left.\frac{\delta n^{0}_{s_1}(\r_1)
  }{\delta m_{s_2}(\r_2)}
  \right|_{m(\r)=m}=n^{0,(2)T}_{s_1 s_2}(\r_1,\r_2)
  \,,
\end{equation}
deriving~(\ref{eq:Xi-1impur}) with respect to $m_{s}(\r)$, we obtain
\begin{subequations}
    \label{eq:np-nm-1impur}
\begin{equation}
  n_{S_1}(\r)=\left[
  1+\alpha_{1} \delta(\r-\R_1)
  \right]
  \left[
    n^{0}_{S_1}(\r)+
    \frac{\alpha_1 n^{0,(2)T}_{S_1 S_1} (\r,\R_1)
    }{1+\alpha_1 n_{S_1}^0(\R_1)}
    \right]
\end{equation}
and
\begin{equation}
  n_{-S_1}(\r)=
    n^{0}_{-S_1}(\r)+
    \frac{\alpha_1 n^{0,(2)T}_{-S_1 S_1} (\r,\R_1)
    }{1+\alpha_1 n_{S_1}^0(\R_1)}
    \,.
\end{equation}
\end{subequations}
When $\Gamma=2$, we recover the results~(\ref{e14}) from
section~\ref{sec:dens-1impur}. Deriving Eqs.~(\ref{eq:np-nm-1impur})
once again with respect to the fugacity, we obtain the truncated
two-body correlation functions
\begin{subequations}
\label{eq:2body-1impur}
\begin{eqnarray}
    n^{(2)T}_{S_1 S_1}(\r_1,\r_2)
    &=&\left[
    1+\alpha_{1} \delta(\r_1-\R_1)
    \right]
  \left[
    1+\alpha_{1} \delta(\r_2-\R_1)
    \right]
  n^{(2)T*}_{S_1 S_1}(\r_1,\r_2)
  \nonumber\\
  \\
  n^{(2)T}_{-S_1 S_1}(\r_1,\r_2)
  &=&
  \left[
    1+\alpha_{1} \delta(\r_2-\R_1)
    \right]
  n^{(2)T*}_{-S_1 S_1}(\r_1,\r_2)\\
  n^{(2)T}_{-S_1, -S_1}(\r_1,\r_2)
  &=&
  n^{(2)T*}_{-S_1, -S_1}(\r_1,\r_2)
\end{eqnarray}
\end{subequations}
with
\begin{eqnarray}
  n^{(2)T*}_{s_1 s_2}(\r_1,\r_2)
  &=&  
  n^{0,(2)T}_{s_1 s_2}(\r_1,\r_2)
  -\frac{\alpha_1^2 n^{0,(2)T}_{s_1 S_1} (\r_1,\R_1)
     n^{0,(2)T}_{S_1 s_2} (\R_1,\r_2)
    }{\left[1+\alpha_1 n_{S_1}^0(\R_1)\right]^2}
  \nonumber\\
  &&+\frac{\alpha_1 n^{0,(3)T}_{s_1 s_2 S_1} (\r_1,\r_2,\R_1)
  }{1+\alpha_1 n_{S_1}^0(\R_1)}
  \,.
  \label{eq:2body-1impur-detail}
\end{eqnarray}
where $n^{0,(3)T}_{s_1 s_2 s_3}$ is the truncated three-body density
function of the unperturbed system. From the Green functions,
Eq.~(\ref{eq:G-1impur-gen}), obtained in the previous section and the
general relations~(\ref{eq:correl-Green}), it can be shown that the
correlation functions obtained in section~\ref{sec:dens-1impur} are
indeed those given by Eqs.~(\ref{eq:2body-1impur})
and~(\ref{eq:2body-1impur-detail}) when $\Gamma=2$.

\subsubsection{Two impurities}
\label{sec:two-impurities-interaction}

Now, let us consider the case of two impurities, located at $\R_1$
and $\R_2$ with adhesivities $\alpha_1$ and $\alpha_2$ and attracting
particles of sign $S_1$ and $S_2$,
respectively. Eq.~(\ref{eq:gp-general}) gives
\begin{equation}
  \label{eq:Xi-2impur}
  \Xi=\Xi_0 \big[1 +\alpha_1 n_{S_1}^{0}(\R_1) +\alpha_2 n_{S_2}^{0}(\R_2)
  +\alpha_1 \alpha_2 n^{0,(2)}_{S_1 S_2}(\R_1,\R_2) \big]
  =\Xi_0\eta
\end{equation}
with $\eta$ defined in Eq.~(\ref{eq:etagen}). The grand potential now
reads 
\begin{eqnarray}\label{e42}
\beta \Omega = \beta\Omega_0-\ln\eta.
\end{eqnarray}
Notice that in the problems for one and two impurities the excess grand
potential is the logarithm of the denominator of the Green functions
$G^1_{s_1s_2}$ and $G^2_{s_1s_2}$ respectively.  It is convenient to
express the grand potential as
\begin{equation}
  \Omega=\Omega_0+\Omega_{\exc}(\alpha_1)+
  \Omega_{\exc}(\alpha_2)+\Omega_{S_1 S_2}(\R_1,\alpha_1;\R_2,\alpha_2)
\end{equation}
with
\begin{equation}\label{e43d}
\beta\Omega_{S_1 S_2}(\R_1,\alpha_1;\R_2,\alpha_2)
=-\ln\left[1+\frac{\alpha_1\alpha_2n^{0,(2)T}_{S_1
      S_2} 
(\mathbf R_1,\mathbf R_2)}
{\big[1+\alpha_1 n_{S_1}^{0}(\R_1)\big]
\big[1+\alpha_2 n_{S_2}^{0}(\R_2)\big]}\right]\,.
\end{equation}
Explicitly, for $\Gamma=2$, 
\begin{eqnarray}\label{e43b}
\beta\Omega_{\pm\pm}(\R_1,\alpha_1;\R_2,\alpha_2)
=-\ln\left[
1-\frac{m^4\alpha_1\alpha_2\big[K_0(m\vert\mathbf R_1-\mathbf R_2\vert)\big]^2}{4\pi^2\big[1+\alpha_1n_0\big]\big[1+\alpha_2n_0\big]}\right]
\end{eqnarray}
and
\begin{eqnarray}\label{e43c}
\beta\Omega_{\pm\mp}(\R_1,\alpha_1;\R_2,\alpha_2)
=-\ln\left[
1+\frac{m^4\alpha_1\alpha_2\big[K_1(m\vert\mathbf R_1-\mathbf R_2\vert)\big]^2}{4\pi^2\big[1+\alpha_1n_0\big]\big[1+\alpha_2n_0\big]}\right].
\end{eqnarray}

A work $\Omega_{\exc}(\alpha_i)$ is required to introduce a single
impurity into the plasma. To introduce two impurities, we require, in
addition to $\Omega_{\exc}(\alpha_1)+\Omega_{\exc}(\alpha_2)$, an
additional work $\Omega_{S_1 S_2}(\R_1,\alpha_1;\R_2,\alpha_2)$. We
can define the effective chemical potential of an impurity with
adhesivity $\alpha$ as $\mu(\alpha)=\Omega_{\exc}(\alpha)$. The term
$\Omega_{S_1 S_2}(\R_1,\alpha_1;\R_2,\alpha_2)$ represents an
effective interaction between the impurities. This contribution is not
a direct interaction between the adsorbing particles but a consequence
of the inhomogeneous charge distribution that they create.


\begin{figure}
\begin{center}
\includegraphics[width=\GraphicsWidth]{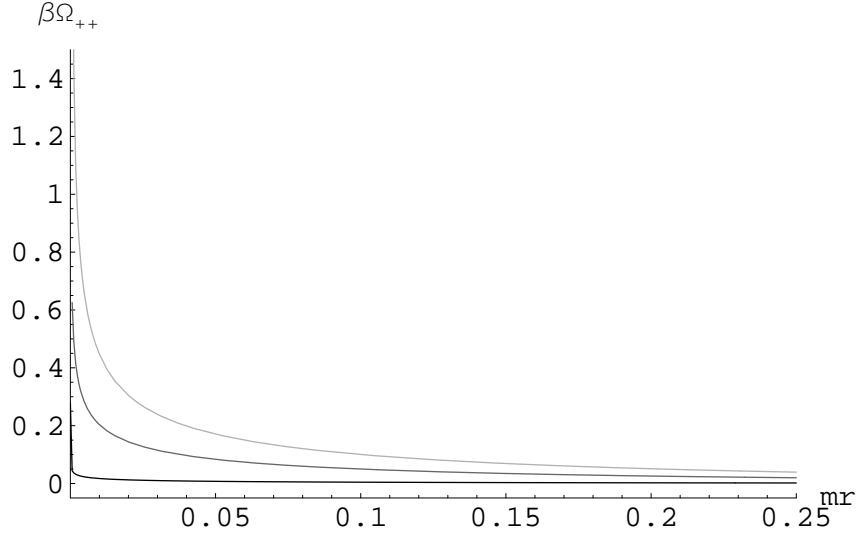}
\caption{\label{fig:Omega-2impur-pp} Effective interaction between two
positive impurities as a function of their distance. The adhesivities
are $m^2\alpha_1=m^2\alpha_2=0.2,1$ and $2$ from darkest to
lightest. The cutoff is $ma=0.01$.}
\end{center}
\end{figure}
\begin{figure}
\begin{center}
\includegraphics[width=\GraphicsWidth]{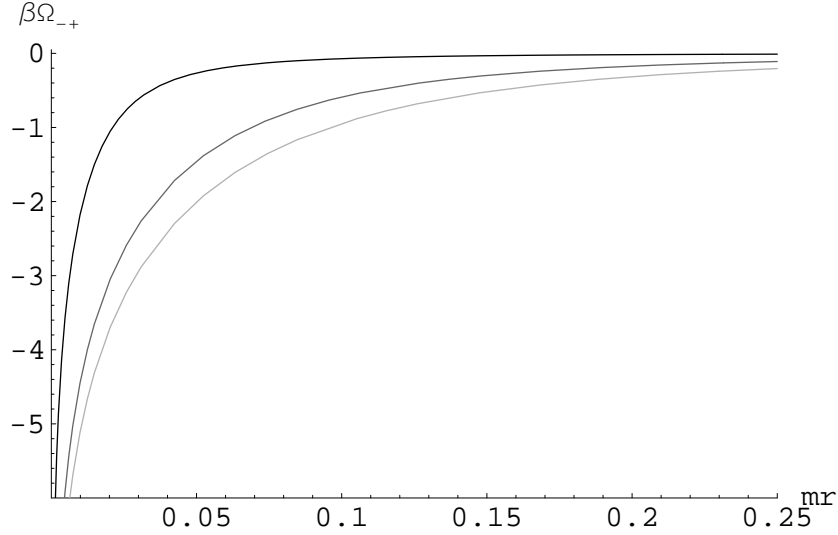}
\caption{\label{fig:Omega-2impur-pm}Effective interaction between a
positive and a negative impurity as a function of their distance. The
values are the same than in figure~\ref{fig:Omega-2impur-pp}.}
\end{center}
\end{figure}


Figures~\ref{fig:Omega-2impur-pp} and~\ref{fig:Omega-2impur-pm} show
the effective interaction between two adsorbing particles. It can be
seen that it is repulsive if the impurities have the same sign and
attractive if their sign is opposite. As expected, the interaction is
important when the impurities are close enough but is negligible for
large distances. It depends on the adhesivity of the two impurities,
indeed, it is larger for large adhesivities.

From Eq.~(\ref{eq:Xi-2impur}), we can obtain the density profiles by
performing a functional derivative with respect to the fugacity. We
obtain
\begin{equation}
\label{eq:n-2impur}
  n_{s}(\r)=
  \begin{cases}
  \left[1+\alpha_1 \delta(\r-\R_1)
    +\alpha_2 \delta(\r-\R_2)
    \right] n^{*}_{s}(\r)
  &\text{if\ }s=S_1=S_2\\
  \left[1+\alpha_1 \delta(\r-\R_1)
    \right] n^{*}_{s}(\r)
    &\text{if\ }s=S_1=-S_2\\
  \left[1+\alpha_2 \delta(\r-\R_2)
    \right] n^{*}_{s}(\r)
    &\text{if\ }s=S_2=-S_1\\
    n^{*}_{s}(\r)
    &\text{if\ }s=-S_1=-S_2\\
  \end{cases}
\end{equation}
with
\begin{eqnarray}
  n_{s}^{*}(\r)
  &=&
  n_{s}^{0}(\r)
  +\frac{1}{\eta}
  \Bigg[
    \alpha_1 n^{0,(2)T}_{s S_1}(\r,\R_1)\left[1+\alpha_2
      n^{0}_{S_2}(\R_2)
      \right]
    \nonumber\\
    &+&
    \alpha_2 n^{0,(2)T}_{s S_2}(\r,\R_2)
    \left[1+\alpha_1 n^{0}_{S_1}(\R_1)\right]
    +
    \alpha_1 \alpha_2 n^{0,(3)T}_{s S_1 S_2}(\r,\R_1,\R_2)
    \Bigg]
  \nonumber\\
  \label{eq:n-2impur-detail}
\end{eqnarray}
With the Green functions~(\ref{eq:Green-gen-2impur}) for two
impurities found in sections~\ref{sec:dens-2impur-pp},
\ref{sec:dens-2impur-pm} and~\ref{sec:dens-2impur-gen} and the general
relation~(\ref{eq:correl-Green}), it is direct to verify that the
density is indeed given by Eqs.~(\ref{eq:n-2impur})
and~(\ref{eq:n-2impur-detail}) when $\Gamma=2$.

We can compute the average number of adsorbed particles by each
impurity $N_{S_i}^{\alpha_i}$, for $i=1,2$. This can be obtained
either from the relation $N_{S_i}^{\alpha_i}=\alpha_{i}
n^{*}_{S_i}(\R_i)$, or from
\begin{equation}
  N_{S_i}^{\alpha_i}=-\alpha_i\frac{\partial (\beta\Omega)}{\partial
  \alpha_i}
  \,.
\end{equation}
We obtain
\begin{equation}
  N_{S_i}^{\alpha_i}=\frac{\alpha_i n^{0}_{S_i}(\R_i)
    +\alpha_1 \alpha_2 n_{S_1 S_2}^{0,(2)}(\R_1,\R_2)}{
    1
    +\alpha_1 n_{S_1}^{0}(\R_1)
    +\alpha_2 n_{S_2}^{0}(\R_2)
    +\alpha_1 \alpha_2 n^{0,(2)}_{S_1 S_2}(\R_1,\R_2)
  }
  \,.
\end{equation}

\section{The plasma with adsorbing lines}
\label{sec:lines}

\subsection{Density and correlations}

\subsubsection{An infinite line}

As a simple model for electrodes with adsorbing sites, we consider now
that a line of impurities is introduced into the plasma. We shall
consider the case where the impurities are very close to each other,
so there is a continuous adsorbing line in the plasma, located at
$x=x_0$. Since the system is translational invariant in the direction
of the line (the $y$ direction), it is better to work in Cartesian
coordinates and confine the plasma in a rectangular box. The fugacity
can be modeled by
\begin{eqnarray}\label{e28}
m_{\pm}(\mathbf r)=m_{\pm}(x)=m\big[1+\bar\alpha_{\pm}\delta(x-x_0)\big].
\end{eqnarray}
The adhesivity $\bar \alpha_{\pm}$ has length dimensions, unlike the
adhesivity $\alpha_\pm$ for point impurities which has area
dimensions. Actually, $\bar \alpha_{\pm}$ is an effective adhesivity
due to the distribution of the adsorbing particles on the line. We
will assume that the line is composed of ``positive'' impurities, so
$\bar\alpha_+=\bar\alpha_1$ and $\bar\alpha_-=0$.

Due to the translational invariance in the $y$ direction, this
problem is easier to solve via Fourier transform. For this purpose we
define
\begin{eqnarray}\label{e29}
G_{s_1s_2}(\mathbf r_1,\mathbf r_2)=\frac{1}{2\pi}
\int_{\mathbb{R}}\widetilde{G}_{s_1s_2}(x_1,x_2,l)e^{il(y_1-y_2)}dl.
\end{eqnarray}\\
By replacing Eq.~(\ref{e29}) and the given fugacities into Eqs.~(\ref{e2}), we
obtain the system
\begin{subequations}
\label{e30}
\begin{eqnarray}\label{e30.1}
m\big[1+\bar\alpha_1\delta(x_1-x_0)\big]\widetilde G_{++}(x_1,x_2,l)
\nonumber\\
+\left(\frac{d}{dx_1}+l\right)\widetilde G_{-+}(x_1,x_2,l)
&=&\delta(x_1-x_2)\\
\Big(\frac{d}{dx_1}-l\Big)\widetilde G_{++}(x_1,x_2,l)
 +m\widetilde G_{-+}(x_1,x_2,l)&=&0\label{e30.2}\\
m\big[1+\bar\alpha_1\delta(x_1-x_0)\big]\widetilde G_{+-}(x_1,x_2,l)
\nonumber\\
+\Big(\frac{d}{dx_1}+l\Big)\widetilde G_{--}(x_1,x_2,l)
&=&0\label{e30.3}\\
\Big(\frac{d}{dx_1}-l\Big)\widetilde G_{+-}(x_1,x_2,l)
 +m\widetilde G_{--}(x_1,x_2,l)&=&\delta(x_1-x_2)\,.
\label{e30.4}
\end{eqnarray}
\end{subequations}

We assume solutions of the form $\widetilde G_{s_1s_2}=\widetilde
G^0_{s_1s_2}+\widetilde G^1_{s_1s_2}$ where $\widetilde G_{s_1s_2}^0$
are the bulk solutions for $\bar\alpha_1=0$. It is easy to show that
these solutions are given by~\cite{uno}
\begin{subequations}
\label{e31}
\begin{eqnarray}
\widetilde G_{\pm\pm}^0(x_1,x_2,l)
&=&\frac{m}{2k}e^{-k\vert x_1-x_2\vert}\\
\widetilde G_{-+}^0(x_1,x_2,l)
&=&\frac{1}{m}\Big(-\frac{d}{dx_1}+l\Big)\widetilde G_{++}^{0}(x_1,x_2,l)\\
\widetilde G_{+-}^0(x_1,x_2,l)
&=&-\frac{1}{m}\Big(\frac{d}{dx_1}+l\Big)\widetilde G_{--}^{0}(x_1,x_2,l)
\end{eqnarray}
\end{subequations}
with $k=\sqrt{m^2+l^2}$. 

The solutions to Eqs.~(\ref{e30}) are
\begin{eqnarray}\label{e32}
\widetilde G_{++}(x_1,x_2,l)
&=&
\left\{\begin{array}{ll}
\frac{m}{2k}e^{-k\vert x_1-x_2\vert}-\frac{m^3\bar\alpha_1}{2k(2k+m^2\bar\alpha_1)}e^{-k\vert 2x_0-x_1-x_2\vert},&x_0<x_1,x_2,\\
& \hspace{-1cm}\text{or\ } x_0>x_1,x_2\\
\frac{m}{2k+m^2\bar\alpha_1}e^{-k\vert x_1-x_2\vert}&
\hspace{-2.5cm}
x_0\;\textrm{between}\; x_1 \; \textrm{and}\; x_2
\end{array}\right.\\
\widetilde G_{--}(x_1,x_2,l)
&=&\left\{\begin{array}{ll}
\frac{m}{2k}e^{-k\vert
  x_1-x_2\vert}+\frac{m\bar\alpha_1(k+l)^2}{2k(2k+m^2\bar\alpha_1)}
e^{k(2x_0-x_1-x_2)}, & x_0<x_1,x_2\\
\frac{m}{2k+m^2\bar\alpha_1}e^{-k\vert x_1-x_2\vert},& 
\hspace{-2.5cm}
x_0\;\textrm{between}\; x_1 \; \textrm{and}\; x_2\\
\frac{m}{2k}e^{-k\vert
  x_1-x_2\vert}
+\frac{m\bar\alpha_1(k-l)^2}{2k(2k+m^2\bar\alpha_1)}
e^{-k(2x_0-x_1-x_2)},& x_0>x_1,x_2
\end{array}\right. 
\label{e32b}
\end{eqnarray} 

The number density profiles $n_{+}$, $n_{-}$, and charge density
profile $\rho$ (in units of $e$) are obtained using
Eqs.~(\ref{eq:correl-Green}). We find
\begin{subequations}
\begin{eqnarray}
  n_{+}(x)&=&\left[1+\bar{\alpha}_1\delta(x-x_0)\right]n^{*}_{+}(x)\\
  n_{-}(x)&=&n_{-}^{*}(x)\\
  \rho(x)&=&n_{+}(x)-n_{-}(x)=\bar{\alpha}_1\delta(x-x_0)n^{*}_{+}(x_0)
  +\rho^{*}(x)
\end{eqnarray}
\end{subequations}
with
\begin{subequations}
  \begin{eqnarray}
    n_{+}^{*}(x)&=&n_{+}^{0}-m^3\bar{\alpha}_1
    \int_{0}^{\infty}
    \frac{dt}{4\pi}
    \frac{e^{-2\sqrt{t^2+1}m|x-x_0|}}{
      \sqrt{t^2+1}(\sqrt{t^2+1}+\tilde{\alpha})}
    \\
    n_{-}^{*}(x)&=&n_{-}^{0}+m^3\bar{\alpha}_1
    \int_{0}^{\infty}
    \frac{dt}{4\pi}
    \frac{(1+2t^2)\,
      e^{-2\sqrt{t^2+1}m|x-x_0|}}{\sqrt{t^2+1}(\sqrt{t^2+1}+\tilde{\alpha})}\\
    \rho^{*}(x)&=&
    -m^{3} \bar{\alpha}_1
    \int_0^{\infty}\frac{dt}{2\pi}
    \frac{\sqrt{t^2+1}
      \, e^{-2\sqrt{t^2+1}m|x-x_0|}}{\sqrt{t^2+1}+\tilde{\alpha}}
  \end{eqnarray}
\end{subequations}
with $\tilde{\alpha}=m\bar{\alpha}_1/2$.

\begin{figure}
\begin{center}
\includegraphics[width=\GraphicsWidth]{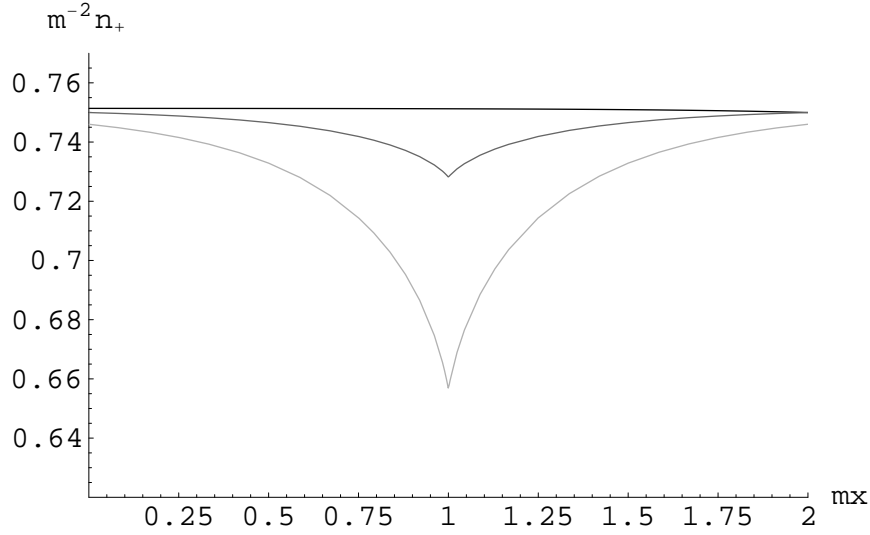}
\caption{\label{fig:plus-1line} Positive density for a line of
``positive'' impurities at $mx=1$. The adhesivities are
$m\bar\alpha_1=0.2,1$ and $3$ from the darkest to the lightest. The
Dirac distribution at $mx=1$ is not plotted. The cutoff is $ma=0.01$.}
\end{center}
\end{figure}

\begin{figure}
\begin{center}
\includegraphics[width=\GraphicsWidth]{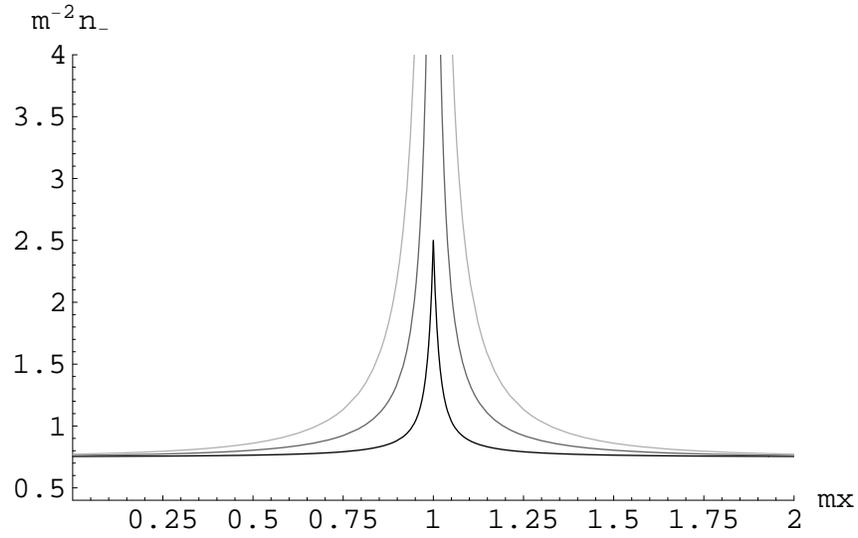}
\caption{\label{fig:minus-1line}Negative density for a line of ``positive'' impurities at $mx=1$. The adhesivities are $m\bar\alpha_1=0.2,1$ and $3$ from the darkest to the lightest. The cutoff is $ma=0.01$.}
\end{center}
\end{figure}

\begin{figure}
\begin{center}
\includegraphics[width=\GraphicsWidth]{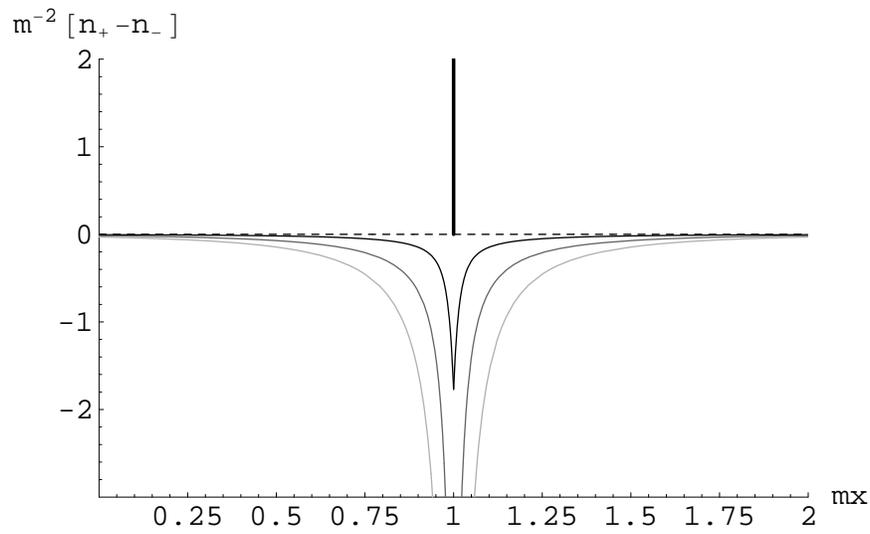}
\caption{\label{fig:charge-dens-1line} Charge density profile for a
line of ``positive'' impurities at $mx=1$. The adhesivities are
$m\bar\alpha_1=0.2,1$ and $3$ from the darkest to the lightest. The
cutoff is $ma=0.01$.}
\end{center}
\end{figure}

Figures~\ref{fig:plus-1line}, \ref{fig:minus-1line}
and~\ref{fig:charge-dens-1line} show the positive and negative density
and the charge density profile for this configuration. The curves
obtained are similar to the ones obtained for point impurities. They
differ in the magnitude as well as in the peaks near the adsorbing
particles. The magnitude of the density at the adsorbing line is
reduced as a consequence of the impurities distribution.


The linear charge density adsorbed by the line, $\sigma$, can be
computed by the relation
\begin{equation}
  \sigma=\bar{\alpha}_1 n^{*}_{+}(x_0)
\end{equation}
or by integrating the opposite of the nonadsorbed charge density
\begin{equation}
  \sigma=-\int_{-\infty}^{\infty} \rho^{*}(x)\,dx
  \,.
\end{equation}
Either way, the result is the same, as expected, 
\begin{subequations}
\begin{eqnarray}
  \sigma &=& m^2\bar{\alpha}_1\int_{0}^{t_{\max}}
  \frac{dt}{2\pi}\frac{1}{\sqrt{t^2+1}+\tilde{\alpha}}
  \label{eq:sigma-1line-int}\\
  &=&
  \bar{\alpha}_1 n_{+}^{0}
  -\frac{m}{\pi}
  \frac{\tilde{\alpha}^2}{\sqrt{\tilde{\alpha}^2-1}}
  \arctanh\frac{\sqrt{\tilde{\alpha}^2-1}}{\tilde{\alpha}}
  \label{eq:sigma-1line-explicit}
\end{eqnarray}
\end{subequations}
where we introduced an ultraviolet cutoff $t_{\max}=e^{-C}/(ma)$ to
obtain a finite result. This choice of the cutoff is done to ensure
the correct result for bulk density~\cite{uno}.

Contrary to the case of a point adsorbing impurity, where the adsorbed
charge reached a maximun value of 1 when $\alpha_1\to\infty$, the
adsorbed line charge density is not bounded when
$\bar{\alpha}_1\to\infty$. Indeed, for large $\bar{\alpha}_1$,
\begin{equation}
  \sigma = \bar{\alpha}_1 n^{0}_{+}
  -\frac{m\tilde{\alpha}}{\pi}
  \left[
    \ln (2\tilde{\alpha})+\frac{1}{2\tilde{\alpha}^2}
    \left[
    \ln(2\tilde{\alpha})-\frac{1}{2}
    \right]
    +O(\tilde{\alpha}^{-4})
    \right]
\end{equation}

It is instructing to compute the electric field $E(x)$ and electric
potential $\phi(x)$ created by the adsorbed line charge density and
its polarization cloud. Integrating Poisson equation we obtain, for
$x>0$ and choosing $x_0=0$,
\begin{equation}
  E(x)=em\tilde{\alpha}\int_0^{\infty}
  \frac{e^{-2\sqrt{t^2+1}mx}\,dt}{\sqrt{t^2+1}+\tilde{\alpha}}
\end{equation}
and
\begin{equation}
  \phi(x)=e\tilde{\alpha}
  \int_0^{\infty}
  \frac{e^{-2\sqrt{t^2+1}mx}\,dt}{t^2+1+\tilde{\alpha}\sqrt{t^2+1}}
\end{equation}
Notice that the electric field at the position of the impurity is
related to the adsorbed linear charge density by $E(0^+)=\pi\sigma$,
and as usual we have the expected discontinuity
$E(0^+)-E(0^-)=2\pi\sigma$. The potential drop from the adsorbed line
to infinity is
\begin{equation}
  \phi(0)-\phi(+\infty)=e\tilde\alpha
  \int_0^{\infty}
  \frac{dt}{t^2+1+\tilde{\alpha}\sqrt{t^2+1}}
  =
  \frac{e\tilde{\alpha}}{\sqrt{\tilde{\alpha}^2-1}}
  \arctanh\frac{\sqrt{\tilde{\alpha}^2-1}}{\tilde{\alpha}}
  \,.
\end{equation}

The correlation functions can be obtained from the Green functions
Eq.~(\ref{e32}) and~(\ref{e32b}), using the relation~(\ref{e3.2}). Let
us simply note that the correlation functions have an exponential decay
both in the $x$ direction (transverse to the adsorbing line) and the
$y$ direction (along the adsorbing line), a behavior similar to the
case without adsorbing line. This is different from the algebraic
decay that the correlation functions show parallel along a hard
wall~\cite{JancoPlaneWall1, JancoPlaneWall2, MartinSumRules}.


\subsubsection{Two infinite lines}

For two parallel infinite lines it is also possible to find analytic
expressions for the Green functions. For two lines located at $x=X_1$
and $x=X_2$, both attracting the positive particles, the fugacities
are 
\begin{eqnarray}
  m_+(x)&=&m\left[
    1+\bar{\alpha}_{1} \delta(x-X_1)+\bar{\alpha}_{2} \delta(x-X_2)
    \right]\\
  m_{-}(x)&=&m
\end{eqnarray}
For one line, located at $x=X_1$, attracting the positive particles,
and a second line, located at $x=X_2$, attracting the negative
particles, the fugacities are
\begin{eqnarray}
  m_+(x)&=&m(1+\bar{\alpha}_{1} \delta(x-X_1))\\
  m_{-}(x)&=&m(1+\bar{\alpha}_{2} \delta(x-X_2))
\end{eqnarray}
Without loss of generality, we suppose that $X_1<X_2$. The resolution
of the linear system~(\ref{e2}) satisfied by the Green functions
follows similar steps as for the previous case of one line.  The
explicit expressions for the Green functions depend on the relative
positions of their arguments with respect to the lines. The complete
expressions can be found in Ref.~\cite{tesis-Ferrero}. Let us focus
here in the expressions obtained for the density profiles.

\begin{figure}
  \begin{center}
    \includegraphics[width=\GraphicsWidth]{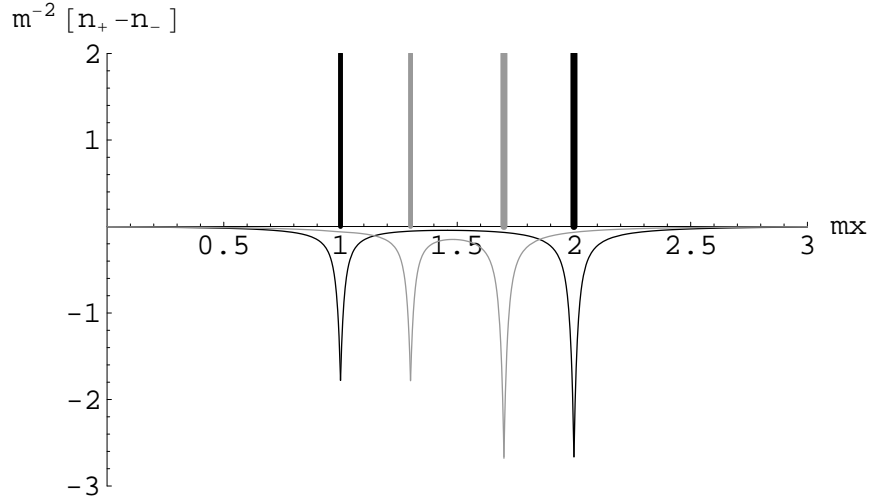}
\caption{\label{fig:charge-2lines} Charge density profile as a
function of the distance. The adhesivities are $m\bar\alpha_1=0.2$ and
$m\bar\alpha_2=0.3$. The position of the lines are $mX_1=1$,
$mX_1=1.3$ and $mX_2=2$, $mX_2=1.7$ from the darkest to the
lightest. The cutoff is $ma=0.01$.}
  \end{center}
\end{figure}

For two lines attracting positive particles, we find the positive,
negative and charge densities
\begin{subequations}
\begin{eqnarray}
  n_{+}(x)&=&\left[1+\bar{\alpha}_1 \delta(x-X_1)
  +\bar{\alpha}_2 \delta(x-X_2)\right]n^{*}_{+}(x)\\
  n_{-}(x)&=&n_{-}^{*}(x)\\
  \rho(x)&=&\left[\bar{\alpha}_1 \delta(x-X_1)
  +\bar{\alpha}_2 \delta(x-X_2)\right]n^{*}_{+}(x) +\rho^{*}(x)
\end{eqnarray}
\end{subequations}
with
\begin{subequations}
  \label{eq:densities-2linepp-a}
\begin{eqnarray}
  n_{+}^{*}(x)&=&n_{+}^{0}
  -\int_{-\infty}^{\infty}
  \frac{m^4 e^{-2kx}}{4\pi k}
  \frac{\bar{\alpha}_1 (2k-m^2\bar{\alpha}_2)e^{2k X_1}
    +\bar{\alpha}_2 (2k+m^2\bar{\alpha}_1)e^{2k X_2}
    }{
    (2k+m^2\bar{\alpha}_1)    (2k+m^2\bar{\alpha}_2)
    -m^4 \bar\alpha_1 \bar\alpha_2 e^{2k(X_1-X_2)}}
  \,dl
  \\
  n_{-}^{*}(x)&=&n_{-}^{0}
  +\int_{-\infty}^{\infty}
  \frac{m^2 (k+l)^2 e^{-2kx}}{4\pi k}
  \frac{\bar{\alpha}_1 (2k-m^2\bar{\alpha}_2)e^{2k X_1}
    +\bar{\alpha}_2 (2k+m^2\bar{\alpha}_1)e^{2k X_2}
    }{
    (2k+m^2\bar{\alpha}_1)    (2k+m^2\bar{\alpha}_2)
    -m^4 \bar\alpha_1 \bar\alpha_2 e^{2k(X_1-X_2)}}
  \,dl
  \nonumber\\
  \\
  \rho^{*}(x)&=&-
  \int_{-\infty}^{+\infty}
  \frac{m^2 k e^{-2kx}}{2\pi}
  \frac{\bar{\alpha}_1 (2k-m^2\bar{\alpha}_2)e^{2k X_1}
    +\bar{\alpha}_2 (2k+m^2\bar{\alpha}_1)e^{2k X_2}
    }{
    (2k+m^2\bar{\alpha}_1)    (2k+m^2\bar{\alpha}_2)
    -m^4 \bar\alpha_1 \bar\alpha_2 e^{2k(X_1-X_2)}}
  \,dl
\end{eqnarray}
\end{subequations}
for $X_1<X_2<x$, with $k=\sqrt{m^2+l^2}$. Between the lines, we find
\begin{subequations}
\label{eq:densities-2linepp-b}
\begin{eqnarray}
  n_{+}^{*}(x)&=&n_{+}^{0}
  \nonumber\\
  &-&\int_{-\infty}^{\infty}
  \frac{m^4}{4\pi k}
  \frac{\bar{\alpha}_1 (2k+m^2\bar{\alpha}_2)e^{2k(X_1-x)}
    +\bar{\alpha}_2 (2k+m^2\bar{\alpha}_1)e^{2k(x-X_2)}
    -2m^2\bar{\alpha}_1 \bar{\alpha}_2 e^{2k(X_1-X_2)}
    }{
    (2k+m^2\bar{\alpha}_1)    (2k+m^2\bar{\alpha}_2)
    -m^4 \bar\alpha_1 \bar\alpha_2 e^{2k(X_1-X_2)}}
  \,dl
  \nonumber\\
  \\
  n_{-}^{*}(x)&=&n_{-}^{0}
  \nonumber\\
  &&
  \hspace{-2cm}
  +\int_{-\infty}^{\infty}
  \frac{m^2 }{4\pi k}
  \frac{\bar{\alpha}_1 (k+l)^2 (2k+m^2\bar{\alpha}_2)e^{2k (X_1-x)}
    +\bar{\alpha}_2 (k-l)^2(2k+m^2\bar{\alpha}_1)e^{2k (x-X_2)}
    +2m^4\bar{\alpha}_1 \bar{\alpha}_2 e^{2k(X_1-X_2)}
    }{
    (2k+m^2\bar{\alpha}_1)    (2k+m^2\bar{\alpha}_2)
    -m^4 \bar\alpha_1 \bar\alpha_2 e^{2k(X_1-X_2)}}
  \,dl
  \nonumber\\
  \\
  \rho^{*}(x)&=&-
  \int_{-\infty}^{+\infty}
  \frac{m^2 k }{2\pi}
  \frac{\bar{\alpha}_1 (2k+m^2\bar{\alpha}_2)e^{2k(X_1-x)}
    +\bar{\alpha}_2 (2k+m^2\bar{\alpha}_1)e^{2k(x-X_2)}
    }{
    (2k+m^2\bar{\alpha}_1)    (2k+m^2\bar{\alpha}_2)
    -m^4 \bar\alpha_1 \bar\alpha_2 e^{2k(X_1-X_2)}}
  \,dl
\end{eqnarray}
\end{subequations}
for $X_1<x<X_2$. It is straightforward to verify that these expressions
reduce to case of one adsorbing line when $X_1\to-\infty$, as it
should. Also, one can verify that for $X_1=X_2$, we recover the density
profiles for one line with adhesivity
$\bar{\alpha}_1+\bar{\alpha}_2$. Figure~\ref{fig:charge-2lines} shows
the charge density profile in the presence of two positive lines.

The linear density of adsorbed particles by each line, $\sigma_1$ and
$\sigma_2$, are obtained from $\sigma_1=\bar{\alpha_1} n_{+}^{*} (X_1)$
and a similar expression for $\sigma_2$. Explicitly,
\begin{eqnarray}
  \sigma_1&=&\bar{\alpha}_1 n_{+}^{0}
  -\frac{m^4\bar{\alpha}_1}{4\pi}
  \int_{-\infty}^{+\infty}
  \frac{\left[\bar{\alpha}_1(2k+m^2\bar{\alpha}_2)
    +\bar{\alpha}_2 (2k-m^2\bar{\alpha_1})
    e^{2k(X_1-X_2)}\right]\,dl
  }{k \left[
      (2k+m^2\bar{\alpha}_1)    (2k+m^2\bar{\alpha}_2)
      -m^4 \bar\alpha_1 \bar\alpha_2 e^{2k(X_1-X_2)}
      \right]}
  \nonumber\\
  &=&
  \frac{m^4}{2\pi}
  \int_{-\infty}^{+\infty}
  \frac{(2k+m^2\bar{\alpha}_2)\bar{\alpha}_1
    -m^2\bar{\alpha}_1 \bar{\alpha}_2 e^{2k(X_1-X_2)}
    }{
      (2k+m^2\bar{\alpha}_1)    (2k+m^2\bar{\alpha}_2)
      -m^4 \bar\alpha_1 \bar\alpha_2 e^{2k(X_1-X_2)}
    }
  \,dl
  \label{eq:sigma1}
\end{eqnarray}
and a similar expression for $\sigma_2$ interchanging $\bar{\alpha}_1$
and $\bar{\alpha}_2$. In the second line of Eq.~(\ref{eq:sigma1}) we
used the formal expression
\begin{equation}
  n_{+}^{0}=\frac{m^2}{4\pi}\int_{-\infty}^{+\infty} \frac{dl}{k}
\end{equation}
for the bulk density (actually, the integral should be cutoff to a
$l_{\max}=e^{-C}/a$ to obtain finite results).  The total adsorbed
charge is found to be
\begin{equation}
  \sigma_1+\sigma_2=
  \frac{m^2}{\pi}
  \int_{-\infty}^{+\infty}
  \frac{(\bar{\alpha}_1+\bar{\alpha}_2)k+m^2 \bar{\alpha}_1
  \bar{\alpha}_2
  \left[1-e^{2k(X_1-X_2)}\right]}{
          (2k+m^2\bar{\alpha}_1)    (2k+m^2\bar{\alpha}_2)
      -m^4 \bar\alpha_1 \bar\alpha_2 e^{2k(X_1-X_2)}
  }
\end{equation}
It is straightforward to verify that 
\begin{equation}
 \sigma_1+\sigma_2=- \int_{-\infty}^{+\infty} \rho^{*}(x)\,dx
\end{equation}
as it should be, since the system is globally neutral.


In the case where line at $X_2$ adsorbs negative particles while the
line at $X_1$ adsorbs the positive ones, we find the following density
profiles
\begin{subequations}
\begin{eqnarray}
  n_{+}(x)&=&\left[1+\bar{\alpha}_1 \delta(x-X_1)\right]
  n^{*}_{+}(x)\\
  n_{-}(x)&=&\left[1+\bar{\alpha}_2 \delta(x-X_2)\right]n_{-}^{*}(x)\\
  \rho(x)&=&\bar{\alpha}_1 \delta(x-X_1) n_{+}^{*}(X_1)-
  \bar{\alpha}_2 \delta(x-X_2)n^{*}_{-}(X_2) +\rho^{*}(x)
\end{eqnarray}
\end{subequations}
with
\begin{subequations}
  \label{eq:densities-2linepm-a}
\begin{eqnarray}
  n_{+}^{*}(x)&=&n_{+}^{0}
  -\int_{-\infty}^{\infty}
  \frac{m^2 e^{-2kx}}{4\pi k}
  \frac{m^2\bar{\alpha}_1 (2k-m^2\bar{\alpha}_2)e^{2k X_1}
    -\bar{\alpha}_2 (k-l)^2(2k+m^2\bar{\alpha}_1)e^{2k X_2}
    }{
    (2k+m^2\bar{\alpha}_1)(2k+m^2\bar{\alpha}_2)
    +m^2 \bar\alpha_1 \bar\alpha_2 (k+l)^2 e^{2k(X_1-X_2)}}
  \,dl
  \nonumber\\
  \\
  n_{-}^{*}(x)&=&n_{-}^{0}
  -\int_{-\infty}^{\infty}
  \frac{m^2 e^{-2kx}}{4\pi k}
  \frac{\bar{\alpha}_1 (k+l)^2 (m^2\bar{\alpha}_2-2k)e^{2k X_1}
    +m^2\bar{\alpha}_2 (2k+m^2\bar{\alpha}_1)e^{2k X_2}
    }{
    (2k+m^2\bar{\alpha}_1)    (2k+m^2\bar{\alpha}_2)
    +m^2 \bar\alpha_1 \bar\alpha_2 (k+l)^2 e^{2k(X_1-X_2)}}
  \,dl
  \nonumber\\
  \\
  \rho^{*}(x)&=&-
  \int_{-\infty}^{+\infty}
  \frac{m^2  e^{-2kx}}{2\pi}
  \frac{\bar{\alpha}_1 (k+l)(2k-m^2\bar{\alpha}_2)e^{2k X_1}
    +\bar{\alpha}_2 (l-k)(2k+m^2\bar{\alpha}_1)e^{2k X_2}
    }{
    (2k+m^2\bar{\alpha}_1)    (2k+m^2\bar{\alpha}_2)
    +m^2 (k+l)^2 \bar\alpha_1 \bar\alpha_2 e^{2k(X_1-X_2)}}
  \,dl
  \nonumber\\
\end{eqnarray}
\end{subequations}
for $X_1<X_2<x$. For $x$ between the lines, $X_1<x<X_2$, we find
\begin{subequations}
  \label{eq:densities-2linepm-b}
\begin{eqnarray}
  n_{+}^{*}(x)&=&n_{+}^{0}
  \nonumber\\
  &&
  \hspace{-2.5cm}
  +\int_{-\infty}^{\infty}
  \frac{m^2}{4\pi k}
  \frac{m^2\bar{\alpha}_1 (-2k-m^2\bar{\alpha}_2)e^{2k(X_1-x)}
    +\bar{\alpha}_2 (k+l)^2 (2k+m^2\bar{\alpha}_1)e^{2k(x-X_2)}
    -2m^2\bar{\alpha}_1 \bar{\alpha}_2 (k+l)^2 e^{2k(X_1-X_2)}
    }{
    (2k+m^2\bar{\alpha}_1)    (2k+m^2\bar{\alpha}_2)
    +m^2 (k+l)^2 \bar\alpha_1 \bar\alpha_2 e^{2k(X_1-X_2)}}
  \,dl
  \nonumber\\
  \\
  n_{-}^{*}(x)&=&n_{-}^{0}
  \nonumber\\
  &&
  \hspace{-2.5cm}
  +\int_{-\infty}^{\infty}
  \frac{m^2}{4\pi k}
  \frac{\bar{\alpha}_1 (k+l)^2 (2k+m^2\bar{\alpha}_2)e^{2k (X_1-x)}
    -m^2\bar{\alpha}_2 (2k+m^2\bar{\alpha}_1)e^{2k (x-X_2)}
    -2m^2\bar{\alpha}_1 \bar{\alpha}_2 (k+l)^2 e^{2k(X_1-X_2)}
    }{
    (2k+m^2\bar{\alpha}_1)    (2k+m^2\bar{\alpha}_2)
    +m^2 (k+l)^2 \bar\alpha_1 \bar\alpha_2 e^{2k(X_1-X_2)}}
  \,dl
  \nonumber\\
  \\
  \rho^{*}(x)&=&-
  \int_{-\infty}^{+\infty}
  \frac{m^2 (k+l) }{2\pi}
  \frac{\bar{\alpha}_1 (2k+m^2\bar{\alpha}_2)e^{2k(X_1-x)}
    -\bar{\alpha}_2 (2k+m^2\bar{\alpha}_1)e^{2k(x-X_2)}
    }{
    (2k+m^2\bar{\alpha}_1)    (2k+m^2\bar{\alpha}_2)
    +m^2 (k+l)^2 \bar\alpha_1 \bar\alpha_2 e^{2k(X_1-X_2)}}
  \,dl
  \,.
\end{eqnarray}
\end{subequations}

The linear charge density adsorbed by each line is now given by
$\sigma_1=\bar{\alpha}_1 n_{+}^{*}(X_1)$ and $\sigma_2=-\bar{\alpha}_2
n_{-}^{*}(X_2)$,
\begin{subequations}
\label{eq:sigma-pm}  
\begin{eqnarray}
  \sigma_1&=&\frac{\alpha_1 m^2}{2\pi}
  \int_{-\infty}^{+\infty}
  \frac{\left[2k+m^2\bar{\alpha}_2 +(k+l)^2 \bar{\alpha}_2
      e^{2k(X_1-X_2)}
      \right]\,dl
  }{
    (2k+m^2\bar{\alpha}_1)    (2k+m^2\bar{\alpha}_2)
    +m^2 (k+l)^2 \bar\alpha_1 \bar\alpha_2 e^{2k(X_1-X_2)}
  } 
  \nonumber\\
  \\
  \sigma_2&=&-\frac{\alpha_2 m^2}{2\pi}
  \int_{-\infty}^{+\infty}
  \frac{\left[2k+m^2\bar{\alpha}_1 +(k+l)^2 \bar{\alpha}_1
      e^{2k(X_1-X_2)}
      \right]\,dl
  }{
    (2k+m^2\bar{\alpha}_1)    (2k+m^2\bar{\alpha}_2)
    +m^2 (k+l)^2 \bar\alpha_1 \bar\alpha_2 e^{2k(X_1-X_2)}
  }     
  \nonumber\\
\end{eqnarray}
\end{subequations}
The total linear charge density adsorbed by the lines is formally
given by
\begin{equation}
  \sigma_1+\sigma_2=(\bar{\alpha}_1-\bar{\alpha}_2)\frac{m^2}{\pi}
  \int_{-\infty}^{+\infty}
  \frac{k\,dl
  }{
    (2k+m^2\bar{\alpha}_1)    (2k+m^2\bar{\alpha}_2)
    +m^2 (k+l)^2 \bar\alpha_1 \bar\alpha_2 e^{2k(X_1-X_2)}
  }
\end{equation}
Once again, it is straightforward to verify that
\begin{equation}
  \sigma_1+\sigma_2=-\int_{-\infty}^{+\infty}
  \rho^{*}(x)\,dx
\end{equation}
as imposed by the global neutrality of the system.

It is interesting to notice that in the two cases considered in this
section, the density profile of particles that are adsorbed by one
line is a continuous function as it crosses the line, while the
density profile of the particles that are not adsorbed by this line is
discontinuous when crossing such line. However, the discontinuity jump
in this density is exponentially small for large separations between
the two lines, it is of order $e^{-2m|X_1-X_2|}$.

\subsection{Grand potential}

In the following subsections, we compute the grand potential of the
plasma with one or two adsorbing lines at $\Gamma=2$, by solving the
eigenvalue problem~(\ref{e6}).

\subsubsection{An infinite line}
\label{sec:gp-1line}

For this case we confine the plasma into a rectangular box of area
$2L_x \times L_y$ and we work in Cartesian coordinates. The boundaries
and the adsorbing line, located at $x=0$, divide the space in four
regions $x<-L_x$, $-L_x<x<0$, $0<x<L_x$ and $x>L_x$ which will be
labeled by (1), (2), (3) and (4).

The eigenvalue system~(\ref{e6}) takes the form
\begin{subequations}
\label{e44}
\begin{eqnarray}
\label{e44.1}
m\big[1+\bar\alpha_1\delta(x)\big]g(\mathbf
r)&=&2\lambda\partial_zf(\mathbf r),
\\ 
mf(\mathbf
r)&=&2\lambda\partial_{\bar z}g(\mathbf r).
\end{eqnarray}
\end{subequations}
for $-L_x\leq x \leq L_x$. Outside that region,
$m_{+}(x)=m_{-}(x)=0$. Thus we conclude that for $x\not\in[-L_x,L_x]$,
$\partial_z f=0$ and $\partial_{\bar{z}}g=0$, that is $f$ is an
anti-analytic function of $z$ and $g$ is an analytic function of $z$.

The translation symmetry on the $y$ axis allows us to assume
$g(\mathbf r)$ of the form $g(x,y)=\tilde{g}(x)e^{il y}$.  The general
solution for the four regions can be written as
\begin{subequations}
\label{e45}
\begin{eqnarray}
g^{(1)}(x,y)&=&A^{(1)}e^{l x+ily}\\
g^{(2)}(x,y)&=&A^{(2)}e^{k_xx+ily}+B^{(2)}e^{-(k_xx-ily)}\\
g^{(3)}(x,y)&=&A^{(3)}e^{k_xx+ily}+B^{(3)}e^{-(k_xx-ly)}\\
g^{(4)}(x,y)&=&B^{(4)}e^{ lx+ily}
\end{eqnarray}
\end{subequations}
and
\begin{subequations}
\label{e46}
\begin{eqnarray}
f^{(1,4)}(x,y)&=&C^{(1,4)} e^{-lx+ily}\\
f^{(2,3)}(x,y)&=&
\frac{\lambda}{m}\Big(\frac{\partial}{\partial
  x}+i\frac{\partial}{\partial y}
\Big)g^{(2,3)}(x,y).
\end{eqnarray}
\end{subequations}
with $k_x=\sqrt{l^2+\frac{m^2}{\lambda^2}}$. 

Since the eigenfunctions must vanish at $x=\pm\infty$, we conclude
that $g^{(1)}=f^{(4)}=0$ for $l<0$ and $f^{(1)}=g^{(4)}=0$ if
$l>0$. The boundary conditions demand that both $g(x,y)$ and $f(x,y)$
must be continuous at $x=\pm L_x$. At $x=0$, $g(x,y)$ is continuous
and $f(x,y)$ is discontinuous due to the Dirac distribution in
Eq.~(\ref{e44.1}):
\begin{equation}
  f(0^{+},y)-f(0^{-},y)=\frac{m\bar{\alpha}_1}{\lambda} g(0,y)
  \,.
\end{equation}

Assuming $l>0$, these boundary conditions can be expressed by the
linear system
\begin{equation}
\left(\begin{array}{cccc}
(k_x-l)e^{-k_xL_x} & -(k_x+l)e^{k_xL_x} & 0 &0\\
0 & 0 & e^{k_xL_x} & e^{-k_xL_x}\\
1 & 1 & -1 & -1\\
m^2\bar\alpha_1 -\lambda^2(k_x-l)& m^2\bar\alpha_1 +\lambda^2(k_x+l)& \lambda^2(k_x-l) & -\lambda^2(k_x+l)
\end{array}\right)
\left(\begin{array}{cccc}
A_2\\
B_2\\
A_3\\
B_3
\end{array}\right)
=0.\label{e47}
\end{equation}
The functions $g(x,y)$ and $f(x,y)$ must not be zero, hence we demand
that the determinant of the latter matrix must vanish. This condition
leads us to the relation
\begin{equation}\label{c4.92}
\cosh(2k_xL_x)\Big[1+\frac{m^2\bar\alpha_1 l{\lambda}^{-2}}
{2k_x^2}\Big]+\sinh(2k_xL_x)\Big[\frac{l}{k_x}+\frac{m^2\bar\alpha_1
\lambda^{-2}}{2k_x}\Big]+\frac{m^2\bar\alpha_1l \lambda^{-2}}{2k_x^2}=0.
\end{equation}
For each value of $l$, there are several possible solutions to
Eq.~(\ref{c4.92}) for the eigenvalue $\lambda$, which we will denote
as $\{\lambda_{l,n}\}_{n}$. For $l<0$, we obtain (\ref{c4.92})
changing $l$ by $-l$. To obtain the grand potential, we can recognize
its relationship with a Weierstrass product~\cite{Whittaker-Watson} as
follows. Let us define the analytic function, for $l\geq0$,
\begin{eqnarray}\label{e49}
h_l(z)&=&\Big(1+\frac{m^2\bar\alpha_1 lz^2}{2(m^2z^2+l^2)}\Big)\cosh{(2\sqrt{m^2z^2+l^2}L_x)}+\frac{m^2\bar\alpha_1 lz^2}{2(m^2z^2+l^2)}\nonumber\\
&+&\Big(\frac{l}{\sqrt{l^2+m^2z^2}}\Big)\Big(1+\frac{m^2\bar\alpha_1 z^2}{2l}\Big)\sinh{(2\sqrt{m^2z^2+l^2}L_x)}.
\end{eqnarray}
We have $h_l(0)=e^{2lL_x}, h'_l(0)=0$ and $h_l(z)=h_l(-z)$. By
construction, the zeros of $h_l$ are precisely the inverse of the
eigenvalues $1/\lambda_{l,n}$. The representation of $h_{l}(z)$ as a
Weierstrass infinite product is therefore
\begin{equation}
h_{l}(z)=h_l(0)\prod_{n}\left(1-{z}{\lambda_{l,n}}\right)
\end{equation}
Then, we notice that the grand potential is simply
\begin{eqnarray}\label{e50}
\beta\Omega= -\sum_l\ln\prod_{n}\left(1+\lambda_{l,n}\right) =
-\frac{L_y}{\pi}\int_0^{\infty} dl \ln\left[e^{-2lL_x}h_l(-1)\right]
\end{eqnarray}
where we replaced the sum over $l$ by an integral in the thermodynamic
limit $L_y\to\infty$. In the thermodynamic limit, we also consider
that $L_x\rightarrow\infty$, and we approximate
$\cosh(2kL_x)\sim\sinh(2kL_x)\sim e^{2kL_x}/2$, with
$k=\sqrt{l^2+m^2}$. The final result is
\begin{eqnarray}\label{c4.94}
\beta\Omega=\beta\Omega_0+\beta\Omega_{\exc}(\bar\alpha_1),
\end{eqnarray}
with 
\begin{eqnarray}\label{e51}
\beta\Omega_{\exc}(\bar\alpha_1)=-\frac{L_y}{\pi}\int_0^{+\infty}
\ln\Big[1+\frac{m^2\bar\alpha_1}{2\sqrt{m^2+l^2}}\Big]dl
\end{eqnarray}
and $\Omega_0$ the grand potential of the unperturbed system for
this geometry,
\begin{equation}\label{e35}
\Omega_0=-2L_x L_y 
p_b+2L_y\gamma+\mathcal{O}(e^{-2mL_x})
\end{equation}
with the bulk pressure $p_b$ and the surface tension near an
impenetrable wall $\gamma$,
\begin{eqnarray}
  \label{eq:bulk-pres}
\beta
p_b&=&\frac{m^2}{2\pi}\Big[\ln{\frac{2}{ma}}-C+\frac{1}{2}\Big]=n_0+\frac{m^2}{4\pi},\\
\beta\gamma&=&m\Big[\frac{1}{4}-\frac{1}{2\pi}\Big]\,.
\end{eqnarray}
Hence we can conclude that the effective chemical potential of a line
of length $L_y$ and adhesivity $\bar\alpha$ is
$\mu(\bar\alpha)=\Omega_{\exc}^\pm(\bar\alpha)$. The
integral~(\ref{e51}) can be computed explicitly. Actually, one must
introduce an ultraviolet cutoff $l_{\max}=e^{-C}/a$ to obtain a finite
result. This choice of the cutoff is done to ensure that the result
for bulk pressure~(\ref{eq:bulk-pres}) is the same as the one obtained
by integrating the bulk
densities~(\ref{eq:bulk-dens})~\cite{uno}. Neglecting terms that vanish
when $a\to0$, we have
\begin{equation}
  \label{eq:Omega-exc-1line}
  \beta\Omega_{\exc}(\bar\alpha)=
  -\frac{m L_y}{\pi}
  \left[
    \tilde\alpha\ln\frac{2 e^{-C}}{ma}-
    \sqrt{\tilde{\alpha}^2-1}\arctanh
    \frac{\sqrt{\tilde{\alpha}^2-1}}{\tilde\alpha}
      -\frac{\pi}{2}+\tilde{\alpha}
    \right]
\end{equation}
with $\tilde\alpha=m\bar\alpha/2$. It is interesting to compare this
result with the surface tension obtained in a similar problem where
the plasma is confined in a strip with adsorbing
boundaries~\cite{siete}. In that problem the adsorbing line was
located at a hard wall boundary. In our notations, the contribution of
both the hard wall and the adsorbing line was~\cite{siete}
\begin{equation}
  \beta \Omega_{\exc}^{\text{adsorbing wall}}
  =-\frac{m L_y}{4\pi}
  \left[
    m\bar\alpha\ln \frac{2e^{-C}}{ma}
    +1-\pi+m\bar\alpha
    +\frac{1-(m\bar\alpha)^2}{\bar\alpha}\ln(1+m\bar\alpha)
    \right]
\end{equation}
We notice that $\Omega_{exc}(\bar\alpha)+L_y\gamma$ of our present problem is
different from $\Omega_{\exc}^{\text{adsorbing wall}}$. This is to be
expected since each situation is different. In our present problem the
adsorbing line is very far from the hard wall boundaries, while in
Ref.~\cite{siete}, the adsorbing line is on the boundary itself. We
notice however that the (divergent) dominant term, when the cutoff $a$
vanish, for our problem is twice the one for the adsorbing boundary
\begin{equation}
  \Omega_{\exc}(\bar\alpha)\sim 2\Omega_{\exc}^{\text{adsorbing wall}}
  \sim \frac{m L_y}{2\pi} \ln(ma)
  \,.
\end{equation}
The factor 2 between both expressions can be understood if we realize
that in our present problem the plasma is on both sides of the
adsorbing line, while in Ref.~\cite{siete} it is only on one side, the
other side is empty.

From Eqs.~(\ref{e51}) and~(\ref{eq:sigma-1line-int}) or from
Eqs.~(\ref{eq:Omega-exc-1line}) and~(\ref{eq:sigma-1line-explicit}),
one can check that the relation between the adsorbed charge density
$\sigma$ and the excess grand potential
\begin{equation}
  \sigma=-\bar{\alpha}_1\frac{\partial (\beta
  \Omega/L_y)}{\partial\bar{\alpha}_1} 
\end{equation}
is satisfied.

\subsubsection{Two infinite lines}
\label{sec:gp-2lines}

Let us now consider the case where there are two infinite parallel
adsorbing lines which attract the same kind of particles, say
positive. The eigenvalue problem which must be solved is
\begin{subequations}
\label{e52}
\begin{eqnarray}
mf(\mathbf r)&=&2\lambda\partial_{\bar{z}}g(\mathbf r)\\
m\big[1+\bar\alpha_1\delta(x-X_1)+\bar\alpha_2(x-X_2)\big]g(\mathbf r)&=&2\lambda\partial_zf(\mathbf r).
\end{eqnarray}
\end{subequations}
We have placed the first line at $x=X_1$ and the second one at
$x=X_2$. The method of solution is similar to the one for one
line. Following similar steps as in the previous section, we find
after some algebra~\cite{tesis-Ferrero}
\begin{eqnarray}\label{e53}
\beta\Omega=\beta\Omega_0+\beta\Omega_{\exc}(\bar\alpha_1)
+\beta\Omega_{\exc}(\bar\alpha_2)+\beta\Omega^{\pm\pm}
\end{eqnarray}
with
\begin{eqnarray}\label{e54}
\beta\Omega^{\pm\pm}=-\frac{L_y}{\pi}\int_0^{+\infty}
\ln\left[1-\frac{m^4\bar\alpha_1\bar\alpha_2e^{-2k|X_1-X_2|}
}{\left(2k+m^2\bar\alpha_1\right)
\left(2k+m^2\bar\alpha_2\right)}\right]dl 
\end{eqnarray}
and $\beta\Omega_{\exc}(\bar\alpha)$ given in (\ref{e51}). We recall
that $k=\sqrt{m^2+l^2}$.

When the lines attract different types of particles the grand potential is
\begin{eqnarray}\label{e55}
\beta\Omega=\beta\Omega_0+\beta\Omega_{\exc}(\bar\alpha_1)
+\beta\Omega_{\exc}(\bar\alpha_2)+\beta\Omega^{\pm\mp}
\end{eqnarray}
with
\begin{eqnarray}\label{e56}
\beta\Omega^{\pm\mp}=-\frac{L_y}{2\pi}\int_{-\infty}^{+\infty}
\ln\left[1+\frac{m^2\bar{\alpha}_1\bar{\alpha}_2 (k+l)^2e^{-2k|X_1-X_2|}
  }{
    \left(2k+m^2\bar\alpha_1\right)
    \left(2k+m^2\bar\alpha_2\right)}\right]dl.
\end{eqnarray}

The effective interaction between two lines (depending on their sign)
is given by the expressions~(\ref{e54}) and~(\ref{e56}).


\begin{figure}
\begin{center}
\includegraphics[width=\GraphicsWidth]{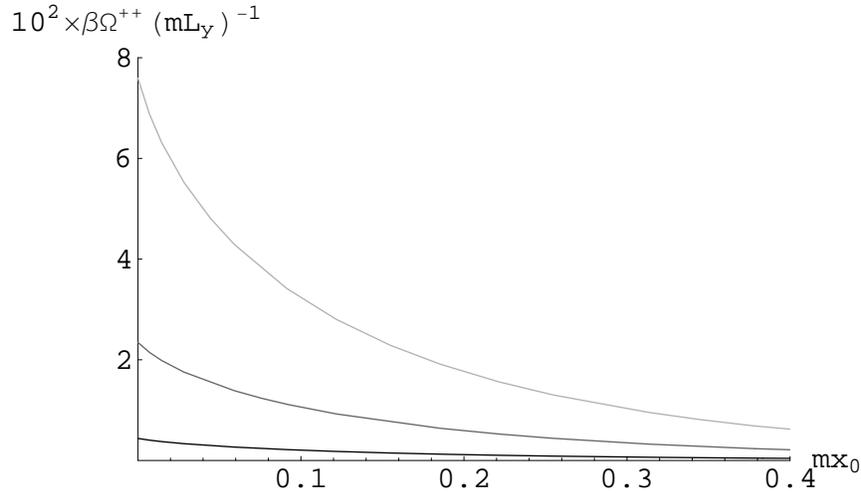}
\caption{\label{fig:inter-2lines-pp} Effective repulsion between two
``positive'' lines as a function of their distance $2x_0=|X_1-X_2|$. The
adhesivities are $m\bar\alpha_1=m\bar\alpha_2=0.2,0.5,1$ from the
darkest to the lightest. The cutoff is $ma=0.01$.}
\end{center}
\end{figure}
\begin{figure}
\begin{center}
\includegraphics[width=\GraphicsWidth]{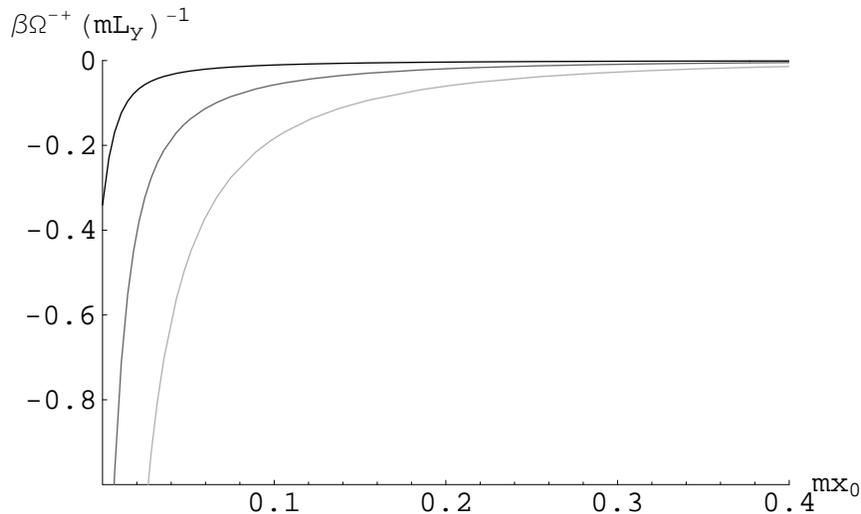}
\caption{\label{fig:inter-2lines-pm} Effective attraction between a
positive and a negative line as a function of their distance
$2x_0=|X_1-X_2|$. The values for the different parameters are the same
as in figure~\ref{fig:inter-2lines-pp}.}
\end{center}
\end{figure}


Figures~\ref{fig:inter-2lines-pp} and~\ref{fig:inter-2lines-pm} show
the effective interaction between two infinite lines. As expected,
like in the problem for point impurities, there is a repulsion if the
sign of the lines is equal and a repulsion if it is different. 

Notice that in both cases, $s=\pm$, $s'=\pm$, in the expression of the
full excess grand potential
$\Omega_{\exc}(\bar{\alpha}_1)+\Omega_{\exc}(\bar{\alpha}_2)+\Omega^{s
s'}$ appears the logarithm of a term which is precisely the denominator
in the Green functions and the density profiles
Eqs.~(\ref{eq:densities-2linepp-a}), (\ref{eq:densities-2linepp-b}),
(\ref{eq:densities-2linepm-a})
and~(\ref{eq:densities-2linepm-b}). Also, one can easily check that
the adsorbed charge on each line, computed in Eqs.~(\ref{eq:sigma1})
and~(\ref{eq:sigma-pm}), can also be
obtained from
\begin{equation}
  \sigma_{i}=-\frac{\beta}{L_y} \alpha_{i}
  \frac{\partial}{\partial\alpha_{i}}
  \left(
    \Omega_{\exc}(\bar{\alpha}_1)+\Omega_{\exc}(\bar{\alpha}_2)+\Omega^{ss'}
    \right)
\end{equation}
for $i=1,2$.

\section{Summary}
\label{sec:summary}

In the present document we analyzed the behavior of the two-component
plasma at $\Gamma=2$ in the presence of one and two point adsorbing
impurities, or in the presence of one and two adsorbing lines.

For point impurities, as shown in Ref.~\cite{cuatro}, the partition
function and correlations of the system can be expressed in terms of
the same quantities for an unperturbed system, without
impurities. Since at $\Gamma=2$, exact results are available for the
partition function and correlations of the unperturbed system, we were
able to obtain exact results for the partition function, the density
profiles and correlation functions of the plasma with one or two point
impurities. We also computed the electric potential created by one
impurity, due to the charge it adsorbs and the polarization cloud that
is formed around it.

As a simple model for electrodes with adsorbing sites, we studied the
properties of the plasma with one or two parallel lines of absorbing
impurities. The general formalism developed in Ref.~\cite{uno} for the
two-component plasma with an external potential is applied to this
case. We obtained exact results for the partition function and density
profiles of the plasma. In both cases we checked that various
relations between the adsorbed charge and the excess grand potential
are satisfied.

\section*{Acknowledgments}

The authors acknowledge partial financial support from Comit\'e de
Investigaciones de la Facultad de Ciencias de la Universidad de los
Andes and from ECOS Nord/COLCIENCIAS.

\end{document}